\documentclass[a4paper,11pt]{article}
\pdfoutput=1 
\usepackage{jheppub} 

\usepackage{amsmath}
\usepackage{rotating}

\usepackage{epsf}
\def\be{\begin{equation}}
\def\ee{\end{equation}}
\def\bea{\begin{eqnarray}}
\def\ba{\begin{array}}
\def\ea{\end{array}}
\def\eea{\end{eqnarray}}\def\nn{\nonumber}

\def\bse{\begin{subequations}}
\def\ese{\end{subequations}}
\def\bt{\begin{tiny}}
\def\et{\end{tiny}}

\title{Renormalization Group Evolution of Neutrino Parameters in Presence
of Seesaw Threshold Effects and Majorana Phases}

\author[a]{Shivani Gupta}
\author[b]{, Sin Kyu Kang}
\author[a]{and C. S. Kim}

\affiliation[a]{Department of Physics and IPAP, Yonsei University, Seoul 120-749, Korea}
\affiliation[b]{School of Liberal Arts, Seoul-Tech, Seoul 139-743, Korea}

\emailAdd{shivani@yonsei.ac.kr}
\emailAdd{skkang@seoultech.ac.kr}
\emailAdd{cskim@yonsei.ac.kr}

\vspace*{1cm}
\abstract
{We examine the renormalization group evolution (RGE) for different mixing scenarios in the presence
 of seesaw threshold effects from high energy scale (GUT) to the low electroweak (EW) scale
 in the Standard Model (SM) and Minimal Supersymmetric Standard Model (MSSM). We consider four
 mixing scenarios namely Tri-Bimaximal Mixing, Bimaximal Mixing, Hexagonal Mixing
and Golden Ratio Mixing which come from different flavor symmetries at the GUT scale.
We find that the Majorana phases play an important role in the RGE running of these mixing
patterns along with the seesaw threshold corrections.
We present a comparative study of the RGE of all these mixing scenarios both with
 and without Majorana CP phases when seesaw threshold corrections are taken into consideration.
 We find that in the absence of these Majorana phases both the RGE running and seesaw
 effects may lead to $\theta_{13}<$ 5$^\circ$ at low energies both in the SM and MSSM.
However, if the Majorana phases are incorporated to the mixing matrix the running can be enhanced both in the SM and MSSM.
Even by incorporating non zero Majorana CP phases in the SM, we do not get $\theta_{13}$ in its present 3$\sigma$ range.
The current values of the two mass squared differences and mixing angles including $\theta_{13}$ can be produced
 in the MSSM case with tan$\beta$ = 10 and non zero Majorana CP phases at low energy.
 We also calculate the order of effective Majorana mass and Jarlskog Invariant for each scenario under consideration.}

\begin{document}
\maketitle
\section{Introduction}

Many flavor symmetries studied in literature \cite{symtbm, symgr} can result
in some particular form of mixing in leptonic sector. The mixing scenarios obtained
by some symmetries lead to the vanishing reactor neutrino mixing angle, $\theta_{13}$.
However, non-zero $\theta_{13}$ has been measured by the reactor experiments \cite{T2K}, it is meaningful
to turn to systematic study of the effects of perturbation on flavor symmetries
or to search for alternative symmetry which gives non zero $\theta_{13}$.
In the flavor basis leptonic mixing matrix is given as
\be
U=
\left(
\ba{ccc}
c_{12}c_{13} & s_{12}c_{13} & s_{13}e^{-i\delta_{CP}} \\
-s_{12}c_{23}-c_{12}s_{23}s_{13}e^{i\delta_{CP}} &
c_{12}c_{23}-s_{12}s_{23}s_{13}e^{i\delta_{CP}} & s_{23}c_{13} \\
s_{12}s_{23}-c_{12}c_{23}s_{13}e^{i\delta_{CP}} &
-c_{12}s_{23}-s_{12}c_{23}s_{13}e^{i\delta_{CP}} & c_{23}c_{13}
\ea
\right)\cdot P.
\ee
Here $c_{ij}=\cos\theta_{ij}, s_{ij}=\sin\theta_{ij}$; $\theta_{ij}$ are the three
mixing angles, $\delta_{CP}$ is the Dirac CP phase. The matrix P=$Diag(1,e^{-i\varphi_1/2},e^{-i\varphi_2/2})$ has
two Majorana CP phases $\varphi_1$ and $\varphi_2$ respectively. The relatively large value of $\theta_{13}$
has also provided an opportunity to measure $\delta_{CP}$ in the lepton mixing matrix.
The Jarlskog rephasing invariant quantity, J$_{CP}$ given as $J_{CP}= c_{12}s_{12}c_{23}s_{23}c_{13}^2s_{13}\sin \delta_{CP}$ \cite{Jarlskog} controls the magnitude of CP violation in neutrino oscillations generated by $\delta_{CP}$.
 Recent global fit analysis for the neutrino parameters is given in \cite{fogli, Tortola, garcia}.
The best fit values along with the 3$\sigma$ constraints on neutrino mass squared differences and mixing angles are given in Table 1.
\begin{table}[t]
\begin{small}
\begin{center}
{\renewcommand{\arraystretch}{1.5}
\begin{tabular}{|c|c|c|c|c}
\hline
Parameter & Best fit & $3\sigma$ Range\\
\hline
\hline
$\Delta m^2_{12}/10^{-5}~\mathrm{eV}^2 $ & 7.50 & 7.00--8.09\\
$\Delta m^2_{13}/10^{-3}~\mathrm{eV}^2 $ & 2.473 & 2.276--2.695 \\
$\theta_{12}^\circ$ & 33.36 & 31.09--35.89\\
$\theta_{13}^\circ$ & 8.66 &  7.19--9.96\\
$\theta_{23}^\circ$ & 40.0, 50.4 & 35.8--54.8            \\
\hline \end{tabular}}
\label{parameter}
\begin{center}
\caption{Experimental constraints on neutrino mass squared differences and mixing angles \cite{garcia}.}
\end{center}
\end{center}
\end{small}
\end{table}
The mixing in neutrino sector is still not completely understood.
We do not know whether the hierarchy of three neutrino masses  is normal ($m_1< m_2 < m_3$) or inverted ($m_3 < m_1 <m_2$).
The CP violating phases are totally unknown at present. The absolute mass scale of neutrinos is still
not known. The possible measurement of effective Majorana mass in neutrinoless double beta decay experiments
will provide an additional constraint on the neutrino mass scale and Majorana CP phases. The effective Majorana mass can be expressed as
\be
M_{ee}=|m_1c_{12}^2c_{13}^2+m_{2}s_{12}^2c_{13}^2e^{-i\varphi_1}+m_3s_{13}^2e^{-i (2\delta_{CP}+\varphi_2)}|.
\ee
The Planck Collaboration \cite{Planck} has given the cosmological constraint on the sum of neutrino masses to be
$\Sigma m_{\nu_i} < $0.23eV at 95$\%$ C.L.. This sum of neutrino masses depend on values chosen for
 the priors and can be in the range $(0.23-0.933)$eV.
The bounds and limits are needed to be tested in the forthcoming observations.

The fact that $\theta_{13}$ is not only non zero but relatively large motivates us to study
how well the flavor symmetries can predict zero value of $\theta_{13}$.
Some of the mixing scenarios from flavor symmetries are Tri-Bimaximal Mixing (TBM) \cite{TBM},
Bimaximal Mixing (BM) \cite{BM}, Hexagonal Mixing (HM) \cite{HM} and Golden Ratio (GR) \cite{symgr, GR1}.
All these mixing scenarios predict the vanishing $\theta_{13}$ and maximal atmospheric mixing angle i.e.
$\theta_{23}=\pi/4$. The solar mixing angle $\theta_{12}$ is different in all four cases. Four different
forms of mixing matrices and the corresponding light neutrino mass matrices considered
here are shown in Table 2.
All the above mixing scenarios can be presented by the matrix form written as
\be
U=
\left(
\ba{ccc}
c_{12} & s_{12} & 0 \\
\frac{-s_{12}}{\sqrt{2}} &
\frac{c_{12}}{\sqrt{2}}& \frac{-1}{\sqrt{2}}\\
\frac{-s_{12}}{\sqrt{2}} &
\frac{c_{12}}{\sqrt{2} }&\frac{1}{\sqrt{2}}
\ea
\right),
\ee
where $\theta_{12}$ is given by $\arcsin(1/\sqrt{3})$ for TBM, $\pi/4$ for BM, $\pi/6$ for HM, and $\tan^{-1}(1/\alpha)$ with $\alpha=(1+\sqrt{5})/2$ for GR.
Mixing angles in these scenarios are determined independent of the neutrino masses.
The mass matrices having such diagonalizing mixing matrix are called mass independent
textures or form diagonalizable textures \cite{formdiag}. There have been some studies earlier on the origin
and effects of perturbations on these mixing scenarios \cite{ourpaper}
in order to accommodate non zero $\theta_{13}$.

\begin{table}
\begin{tiny}
\begin{center}
{\renewcommand{\arraystretch}{2.0}
\begin{tabular}{|c|c|c|p{3.3cm}|}
\hline
Mixing& U & $M_{\nu}$ &A, B, C \\
\hline
\hline
TBM & $\begin{tiny}\left(
 \ba {ccc}
 \frac{2}{\sqrt{6}}& \frac{1}{\sqrt{3}}  & 0 \\
\frac{-1}{\sqrt{6}} &\frac{1}{\sqrt{3}}  & \frac{-1}{\sqrt{2}} \\
\frac{-1}{\sqrt{6}}& \frac{1}{\sqrt{3}}  & \frac{1}{\sqrt{2}}
\ea
\right)\end{tiny}$ & $\begin{tiny}\left(
\ba{ccc}
A &B &B \\
.. & \frac{1}{2}(A+B+C) &\frac{1}{2}(A+B-C) \\
.. & . &\frac{1}{2}(A+B+C)
\ea \right)\end{tiny}$ & $A=
\frac{1}{3}(2m_1+m_2 e^{-i\varphi_1})$,
$B=\frac{1}{3}(m_2 e^{-i\varphi_1 }-m_1)$,
$C=m_3 e^{-i\varphi_2}$\\


BM & $\begin{tiny}\left(
\ba {ccc}
 \frac{1}{\sqrt{2}}& \frac{1}{\sqrt{2}}  & 0 \\
\frac{-1}{2}& \frac{1}{2} & \frac{-1}{\sqrt{2}} \\
\frac{-1}{2}& \frac{1}{2} & \frac{1}{\sqrt{2}}
\ea
\right)\end{tiny}$ &$\left(
\ba{ccc}
A &~~ B & B \\
.. &~ C &~ A-C \\
.. & . & C
\ea \right)$& $A=
 \frac{1}{2}(m_1+m_2 e^{-i\varphi_1})$,
$B=\frac{1}{2\sqrt{2}}(m_2 e^{-i\varphi_1} -m_1)$,
$C=\frac{1}{4}(m_1+m_2 e^{-i\varphi_1}+2 m_3e^{-i\varphi_2})$\\


HM & $\begin{tiny}\left(
\ba {ccc}
 \frac{\sqrt{3}}{2}& \frac{1}{2}  & 0 \\
\frac{-1}{2\sqrt{2}}& \frac{\sqrt{3}}{2\sqrt{2}} & \frac{-1}{\sqrt{2}} \\
\frac{-1}{2\sqrt{2}}& \frac{\sqrt{3}}{2\sqrt{2}} & \frac{1}{\sqrt{2}}
\ea
\right)\end{tiny}$ & $\begin{tiny}\left(
\ba{ccc}
A &~~ B & B \\
.. &\frac{1}{2}(A+\sqrt{\frac{8}{3}}B+C) &\frac{1}{2}(A+\sqrt{\frac{8}{3}}B-C) \\
.. & . & \frac{1}{2}(A+\sqrt{\frac{8}{3}}B+C)
\ea \right)\end{tiny}$ & $A=\frac{1}{4}(3m_1+m_2 e^{-i\varphi_1})$,
$B=\frac{1}{4}\sqrt{\frac{3}{2}}(m_2 e^{-i\varphi_1} -m_1)$,
$C=m_3e^{-i\varphi_2}$\\


GR & $\begin{tiny}\left(
\ba {ccc}
 \frac{\alpha}{\sqrt{1+\alpha^2}}& \frac{1}{\sqrt{1+\alpha^2}}  & 0 \\
\frac{-1}{\sqrt{2(1+\alpha^2)}}& \frac{\alpha}{\sqrt{2(1+\alpha^2)}} & \frac{-1}{\sqrt{2}} \\
\frac{-1}{\sqrt{2(1+\alpha^2)}}&\frac{\alpha}{\sqrt{2(1+\alpha^2)}} & \frac{1}{\sqrt{2}}
\ea
\right)\end{tiny}$  & $\begin{tiny}\left(
\ba{ccc}
A &~~ B & B \\
.. &~ C &~ A+\sqrt{2}B-C \\
.. & . & C
\ea \right)\end{tiny}$ & $A= \frac{1}{(1+\alpha^2)}(m_1\alpha^2+m_2 e^{-i\varphi_1})$,
$B=\frac{1}{\sqrt{2}(1+\alpha^2)}(m_2 e^{-i\varphi_1} -m_1)\alpha$,
$C=\frac{1}{2(1+\alpha^2)}(m_1+m_2 e^{-i\varphi_1}\alpha^2)+\frac{1}{2}m_3e^{-i\varphi_2})$ \\

\hline \end{tabular}}
\label{form}
\begin{center}
\caption{The mixing matrices $U$ and their corresponding light neutrino mass matrices, $M_{\nu}$.
For GR, $\alpha=(1+\sqrt{5})/2$ as given in the text.}
\end{center}
\end{center}
\end{tiny}
\end{table}

Another attractive possibility is that those flavor symmetries are present at very
high scale, namely, grand unified scale ($\Lambda_{GUT}\sim$10$^{16}$GeV).
It has been found earlier in \cite{RGTBM, CP, review} that corrections from the renormalization group evolution (RGE) can significantly
affect neutrino mixing angles, CP phases and mass splittings and thus, they should not be neglected in the models with flavor symmetries imposed at high energy scale.

In the framework of type I seesaw with three heavy right handed neutrinos \cite{Seesaw}, we study the radiative corrections to the masses and mixing angles of neutrinos in the charged lepton basis
by the RGE \cite{RGE, Antush, Antush2} from $\Lambda_{\rm GUT}$ to $\Lambda_{\rm EW}(\sim$10$^2$GeV), in addition to the seesaw threshold corrections \cite{Antusch:2005gp, xing, bergstrom}. Threshold corrections occur by subsequently
integrating out heavy right handed Majorana masses at the respective seesaw scales both in the SM and MSSM.
We assume that all the above mentioned specific mixing matrices are realized at GUT scale,
and the corresponding light neutrino mass matrices, $M_{\nu}$, are given in terms of three masses as shown in Table 2.
The heavy right handed neutrino mass matrices can be determined by inverting the seesaw formula. We first take the
 general form of the neutrino Dirac Yukawa matrix $Y_{\nu}$  and then pick its specific form that leads to the specific mixing pattern by scanning the parameter space.
Below the seesaw threshold scales the RGE behavior is described by the effective theories which
are governed by the effective mass operators. However, above all the seesaw threshold scales, we have to
consider the full theory. The interplay of the heavy and the
light sectors can modify the RGE effects, further on top of what were in the effective theory.
In Ref. \cite{xing}, the authors have studied RGE evolution of neutrino mixing angles and CP phases for some mixing
 scenarios at high scale by incorporating seesaw threshold effects and concluded that two of the considered mixing
 scenarios can lead to at most $\theta_{13}\sim$ 5$^\circ$ at $\Lambda_{\rm EW}$.
However, they do not fully consider the effects of the Majorana phases in the RGE evolution.
Comparatively studying the RGE in absence and presence of Majorana phases, we find that these phases can give significant contributions in the running of neutrino flavor mixing angles.
We have checked that turning off some CP phases in the present study reproduces the results almost similar to \cite{xing}.
We have found that the measured reactor mixing angle, $\theta_{13}$, up to $3 \sigma$ C.L. can be achieved {\it only when} the Majorana phases are fully incorporated in the RGE from the GUT scale to
electroweak scale.
In fact, in order to realize such mixing scenarios, one may need to add additional Higgs bosons. The existence of the extra particles may affect the RG running, but those effects are highly model dependent. In particular, the RGEs for dimension five neutrino mass operator in the multi-Higgs doublet models are derived and their running has been performed in \cite{Grimus}.
In this work, we assume that  the contributions of extra particles to the RG running
are negligibly small, which can be achieved by taking couplings to be small.

The paper is organized as follows.
In section 2, we discuss the specific forms of neutrino mass and mixing matrices for different mixing scenarios at the GUT scale.
In section 3, the RGE equations governing at various energy scales in addition to the seesaw
threshold effects are presented.
Section 4 gives the numerical results of our study both in the SM and the MSSM, respectively.
We summarize our results in the last section.

\section{Lepton mixing matrices at the GUT scale}

In the basis where the charged lepton mass matrix is real and diagonal, the effective light neutrino mass matrix, $M_{\nu}$, is in general given as
\be
\label{mnu}
M_{\nu} = U^*P^* M_{\nu}^{diag}P^\dagger U^\dagger.
\ee
Here $U$ has one of the forms given in Table 2. P is the phase matrix having two Majorana
phases given as Diag($1, e^{-i\varphi_1/2},e^{-i\varphi_2/2}$) and $M_{\nu}^{diag}$= Diag$(m_1, m_2, m_3)$.
The different form of the corresponding light neutrino mass matrices are given in Table 2.
Following exactly the same procedure as given in Ref. \cite{xing},
the Yukawa coupling matrix, $Y_{\nu}$,  is taken to be of the form
\be
Y_{\nu}=y_{\nu}\cdot R \cdot Diag(r_1,r_2,1).
\ee
The three real, positive and dimensionless parameters $y_{\nu}$, $r_1$ and $r_2$ characterize the
hierarchy of $Y_{\nu}$ and R is given as
\be
R=R_{23}(\theta_2)\cdot R_{13}(\theta_3 e^{-i\delta})\cdot R_{12}(\theta_1),
\ee
where R$_{ij}$ are the rotation matrices in the $ij$th plane. The three mixing angles ($\theta_1, \theta_2, \theta_3$)
 and a phase $\delta$ are free parameters varied randomly. Thus, $Y_{\nu}$ comprises of 7 free parameters,
 three eigenvalues, three mixing angles and $\delta$.
The parameters $y_{\nu}$, $r_1$ and $r_2$ are small $\leq \mathcal{O}(1)$.
Thus, the effective RGEs between GUT scale and seesaw scales depend on  $Y_{\nu}$, $Y_l$ and $M_{R}$.
$M_{R}$ can be determined by inverting seesaw formula given as
\be \label{Mr1}
M_{R}=- Y_{\nu} M_{\nu}^{-1}Y_{\nu}^T.
\ee
Transferring to the basis where $M_R$ is diagonal,
\be \label{Mr2}
U_{R}^TM_RU_R=Diag(M_{R_1},M_{R_2},M_{R_3}).
\ee
$Y_{\nu}$ gets simultaneously transformed as $Y_{\nu}U_R^*$. Since $M_R$ in our analysis is hierarchical i.e.
$M_{R_1}<M_{R_2}<M_{R_3}$ we consider the seesaw threshold effects
which arise due to sequential decoupling of these fields at respective scales.
For the normal hierarchical spectrum the lowest neutrino mass, $m_1$ is a free parameter.
The other two masses $m_2$ and $m_3$ are determined by the relation $m_2=\sqrt{m_1^2+\Delta m_{12}^2}$ and $m_3=\sqrt{m_1^2+\Delta m_{13}^2}$ where $\Delta m_{12}^2$ and $\Delta m_{13}^2$ are the solar and the atmospheric mass squared differences, respectively. We present the numerical analysis for normal hierarchical spectrum where $m_1$ is the smallest mass. Since the running of the mixing angles is inversely proportional to masses there can be more corrections to those angles for the quasidegenerate and inverted mass spectrum in these mixing scenarios \cite{Antush2}. Yet, we focus on whether the measurements on $\theta_{13}$ can be achieved by RG running in the normal hierarchical spectrum,
within the most conservative scenario.


\section{RGE equations in the presence of seesaw threshold effects}

Extended by three right handed neutrinos, the leptonic Yukawa terms of the Lagrangian in the SM can be written as
\be
-\mathcal L_{({\rm SM})}= \bar l_l H Y_l l_R +\bar l_l \tilde H Y_{\nu} \nu_R +\frac{1}{2}\bar {\nu^{c}}_R M_R \nu_R +h.c..
\ee
 For the MSSM it is
\be
-\mathcal L_{({\rm MSSM})}= \bar l_l H_1 Y_l l_R +\bar l_l \tilde H_2 Y_{\nu} \nu_R +\frac{1}{2}\bar {\nu^{c}}_R M_R \nu_R +h.c.,
\ee
where $H$ ($\tilde H$=$i \sigma^2 H^*$) is the SM Higgs doublet ($H_1, H_2$ for MSSM), $l_l$, $e_R$, $\nu_R$ are
the lepton $SU(2)_L$ doublet, right handed charged leptons and right handed neutrinos, respectively.
The current neutrino mixing angles and mass squared differences
are determined from the neutrino oscillation experiments at the low energy scale.
The seesaw threshold corrections can be quite significant
at the seesaw scales as the heavy singlets can be nondegenerate i.e. $M_{R_1}<M_{R_2}<M_{R_3}$.
In the flavor basis the effective light neutrino mass matrix, $M_{\nu}$, above the highest seesaw scale is given to be
\be
M_{\nu}(\mu)= -\frac{\kappa (\mu)v^2}{4},
\ee
here $v$ =246 GeV in the SM and (246 GeV)$\cdot$sin$\beta$ in the MSSM, $\mu$
 is the renormalization scale and $\kappa$ is the effective coupling matrix given as \\
\be
\kappa (\mu) = 2 Y_{\nu}^T(\mu)M_R^{-1}(\mu)Y_{\nu}(\mu).
\ee
For the running of neutrino parameters above the seesaw scales we use the formulae given in \cite{Antusch:2005gp}.
The radiative corrections from the GUT to $M_{R_3}$ comprises of running of the Yukawa
 couplings $Y_{\nu}$, $M_R$ and $Y_l$. We work in the basis where charged lepton is diagonal.
 In the course of running there are additional contributions generated from off-diagonal entries
 of $Y_l^{\dagger}Y_l$, which are taken into consideration while calculating the total mixing matrix as $U_l^{\dagger}U_{\nu}$.
The effective  operator at the heaviest scale, $M_{R_3}$ is given by the matching condition
\be \label{matching}
\kappa^{(3)} = 2 Y_{\nu}^{T}M_{R_3}^{-1}Y_{\nu},
\ee
in the basis where $M_{R}$ is diagonal. The effective neutrino mass matrix at the scale below $M_{R_3}$ now constitutes of two parts
\be \label{mnu3}
M_{\nu}=-\frac{v^2}{4}[\kappa^{(3)}+ 2Y_{\nu}^{T(3)}M_R^{-1(3)}Y_{\nu}^{(3)}],
\ee
where $Y_{\nu}^{(3)}$ is 2$\times$3 and $M_R^{(3)}$ is the 2$\times$2
mass matrix remained after decoupling $M_{R_3}$.
Following the same procedure at $\mu=M_{R_2}$ the Yukawa coupling matrix is further reduced to 1$\times$3 matrix.
The one loop RGE for $\kappa^{(1)}$  after decoupling all the three heavy right handed fields \cite{Antusch:2005gp} is given as
\be
 16\pi^2 \frac{d\kappa^{(1)}}{dt}=  (C_{\nu}^l (Y_l^{\dagger}Y_l)^T) \kappa^{(1)} + \kappa^{(1)}(C_{\nu}^l (Y_l^{\dagger}Y_l)) +\alpha \kappa^{(1)},
\ee
where parameter $\alpha$ is explicitly given by
\be
\label{alp}
\alpha_{\rm SM}= 2\ Tr(3Y_u^{\dagger}Y_u+3Y_d^{\dagger}Y_d + Y_l^{\dagger}Y_l) - 3g_2^2 + \lambda ,
\ee
\be \nn
\alpha_{\rm MSSM}= 2\ Tr(3Y_u^{\dagger}Y_u)-\frac{6}{5}g_1^2-6g_2^2.
\ee
The effective neutrino mass matrix obtained from $\kappa^{(1)}$ at the EW scale is diagonalized to obtain the neutrino mixing angles, CP violating phases and mass squared differences.

The neutrino mass matrices at two different scales $\Lambda_{GUT}$ and $\Lambda_{EW}$ are homogeneously related as \cite{Lola, pokorski}
\be
M_{\nu}^{\Lambda_{EW}}= I_KI^T_\kappa M_{\nu}^{\Lambda_{GUT}}I_\kappa,
\ee
here $I_K$ is the scale factor common to all elements of $M_{\nu}^{\Lambda_{EW}}$. The matrix $I_{\kappa}$ is given as
\be
I_{\kappa}=Diag(\sqrt{I_e},\sqrt{I_{\mu}},\sqrt{I_{\tau}}).
\ee
In the presence of seesaw threshold corrections \cite{khan} we have \\
\begin{center} $\sqrt{I_j}=Exp{(-\frac{1}{16\pi^2}\int[3(Y_j^{\dagger} Y_j)-(Y_{\nu_j}^{\dagger}Y_{\nu_j})]dt)}=e^{-\Delta_j}$,\end{center}
j=$e$, $\mu$ and $\tau$.
From this relation when  $Y_{\tau}\sim 0.01$ and $Y_{\nu_\tau}\sim 0.3$, the magnitude of $\Delta_{\tau}$
can be of the order of 10$^{-3}$ in the SM (10$^{-3}(1+\tan^2 \beta)$ in the MSSM) from 10$^{12}$GeV to 10$^2$GeV.
It is worthwhile to notice that in presence of threshold effects, the magnitudes of $\Delta_e$
and $\Delta_{\mu}$ can be comparable to $\Delta_{\tau}$ for large values of $Y_{\nu_e}$ and $Y_{\nu_\mu}$.
Thus, $\Delta_e$ and $\Delta_{\mu}$ terms in the presence of threshold effects can play an important role in
enhancing RGE corrections.
 In the absence of threshold effects when $Y_{\nu_\tau}$ is absent, $\Delta_\tau$ is of the order of $10^{-5}$ and $\Delta_e$, $\Delta_{\mu}$ contributions are very small and thus can be neglected.

Below the seesaw scales where all the heavy right handed fields are integrated out,
the running in the SM is mostly governed by $Y_{\tau}\sim \sqrt{2}m_{\tau}/v \approx \mathcal{O} (10^{-2})$
and $Y_{\tau}\sim \sqrt{2}m_{\tau}/{(v\cos\beta)}$ in the MSSM.
There are no significant corrections to mixing angles in the SM even for quasidegenerate spectrum of masses in this region.
In the MSSM case, however, there can be significant corrections when tan$\beta$ is large.
The analytic expressions of the RGE of masses, mixing angles and CP phases below the
seesaw scales are given in \cite{Antush2} and the expressions for running above the
seesaw scales are given in \cite{Antusch:2005gp, bergstrom} in detail.

\section{Numerical results}

We begin at $\Lambda_{\rm GUT}$ by varying the three angles and phase of $Y_{\nu}$
along with two Majorana phases in the range of 0--2$\pi$. Three hierarchical parameters of $Y_{\nu}$ and $m_1$
are also randomly varied.
We note that the Dirac phase $\delta_{CP}$ is not well defined when $\theta_{13}$ becomes zero at the GUT scale.
In Ref. \cite{Antush2} the analytical continuation condition is derived that assures $\frac{d \delta_{CP}}{dt}$ is finite and
running of $\delta_{CP}$ is extended continuously even when $\theta_{13}$ approaches to zero.
Following \cite{Antush2}, we can avoid the divergence happened in the running of $\delta_{CP}$.
Since there are lots of free parameters, it is hard to obtain full parameter space in consistent with
experimental data up to $3 \sigma$ C.L..
Instead, we present some sets of input parameter space which lead to maximally allowed value of $\theta_{13}$ achieved at low scale, while
the other mixing angles and mass squared differences are simultaneously in the ranges of measured values
up to 3$\sigma$.
\begin{table}[t]
\begin{small}
\label{resultstbm1}
\begin{center}
\begin{tabular}{c|cc|c|cc}
\hline
\hline
SM Input  & $\varphi_1,\varphi_2=0$  & $\varphi_1,\varphi_2\neq0$ & SM Output &$\varphi_1,\varphi_2=0$ &  $\varphi_1,\varphi_2\neq0$  \\
\hline
$r_1$ & 0.57$\times 10^{-3} $ & 0.83$\times 10^{-3}$& $M_{R_1}$ (GeV) & 2.6$\times$10$^4$& 4$\times$10$^3$ \\
$r_2$ & 0.6 & 0.47&$M_{R_2}$ (GeV)& 2.1 $\times$10$^9$& 8.2 $\times$10$^8$\\
$\delta$ & 14.7$^\circ $ & 28.36$^\circ$&$M_{R_3}$ (GeV)&1.7 $\times$ 10$^{11} $&2.5 $\times$ 10$^{9}$\\
$y_{\nu}$& 0.5 &  0.35& -& -&-\\
$\theta_1$ & 216$^\circ$  & 194.2$^\circ$ &$\theta_{12}$ &35.1$^\circ$ &33.8$^\circ$  \\
$\theta_2$ &168$^\circ$ & 81.36$^\circ$ &$\theta_{23}$	& 44.1$^\circ$&43.9$^\circ$ \\
$\theta_3$ & 80.8$^\circ$ & 350.6$^\circ$  &$\theta_{13}$& 1.6$^\circ$&4.3$^\circ$\\
$ m_1$ (eV)& 3.6$\times 10^{-4}$& 0.023 &$ m_1$ (eV)&2.94$\times 10^{-4} $&0.021\\
$\Delta m^2_{12}$(eV$^2$) & 8.2 $\times$ 10$^{-5}$  & 8.0 $\times$ 10$^{-5}$ &$\Delta m^2_{12}$(eV$^2$)& 7.63 $\times$ 10$^{-5}$&7.86 $\times$ 10$^{-5}$\\
$\Delta m^2_{13}$(eV$^2$)& 3.7 $\times$ 10$^{-3}$& 3.3 $\times$ 10$^{-3} $&$\Delta m^2_{13}$(eV$^2$) &2.6$\times$ 10$^{-3}$ &2.48$\times$ 10$^{-3}$ \\
$\varphi_1$ & 0$^\circ$ & 281.7$^\circ$&$\varphi_1$&124$^\circ$ & 351$^\circ$\\
$\varphi_2$& 0$^\circ$& 140.4$^\circ$&$\varphi_2$&138$^\circ$ & 227.5$^\circ$\\
--&--&--&$J_{CP}$& 5.8 $\times$10$^{-3}$& 0.016 \\
--&--& --&$M_{ee} (eV)$&2.85$\times$10$^{-3}$ & 0.02 \\
\hline
\end{tabular}
\end{center}
\begin{center}
\caption{Numerical values of input and output parameters radiatively generated in the SM for TBM mixing
for zero and non zero Majorana phases at $\Lambda_{\rm GUT}$ = 2$\times 10^{16}$GeV.
The input values for neutrino mixing angles at the GUT scale are $\theta_{13}$= 0$^\circ$, $\theta_{23}$= 45$^\circ$ and $\theta_{12}$= 35.3$^\circ$.}
\end{center}
\end{small}
\end{table}
\begin{figure}[h]
\begin{center}
\label{figtbmsm}
\includegraphics[width=0.42\textwidth]{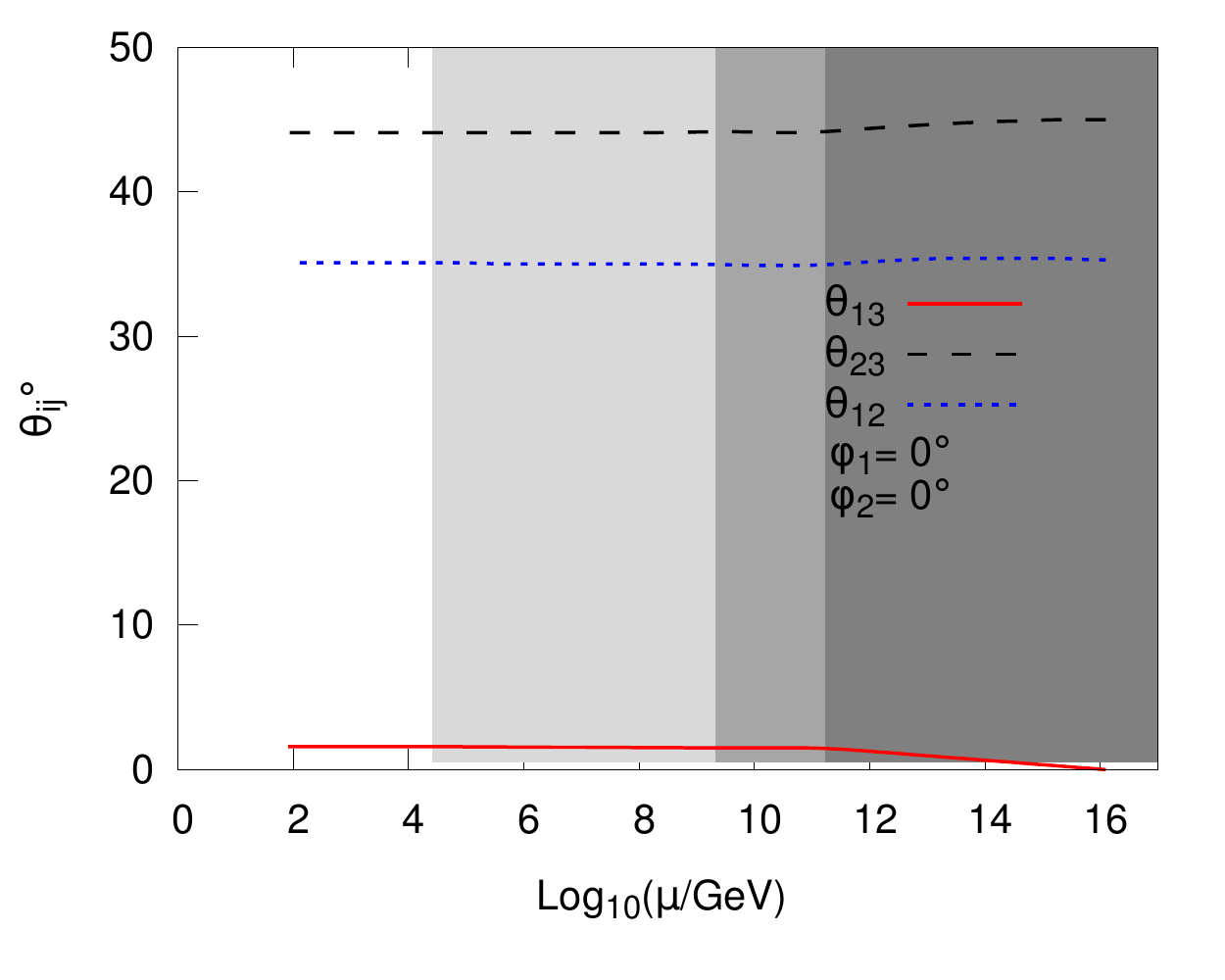}
\includegraphics[width=0.42\textwidth]{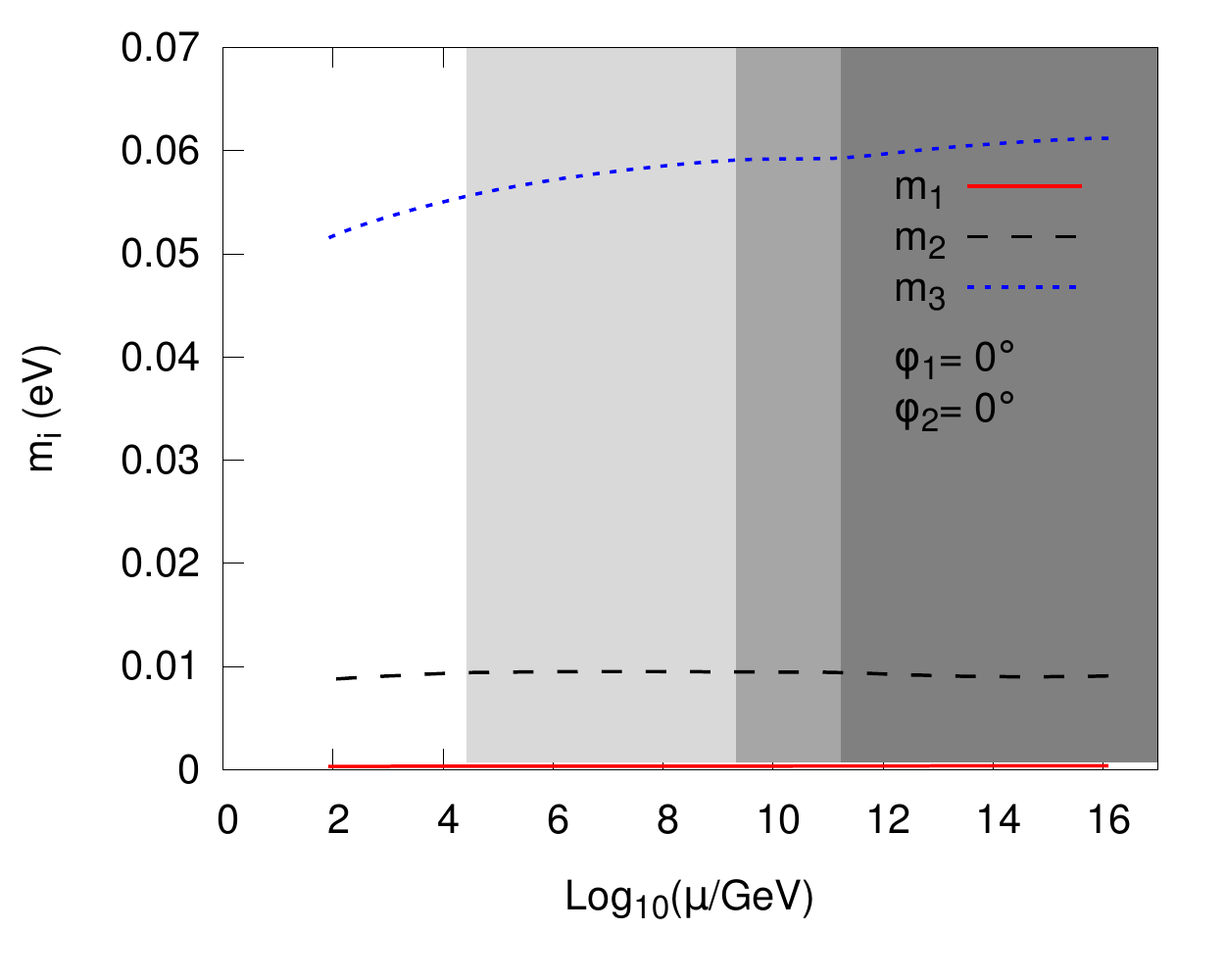}
\includegraphics[width=0.42\textwidth]{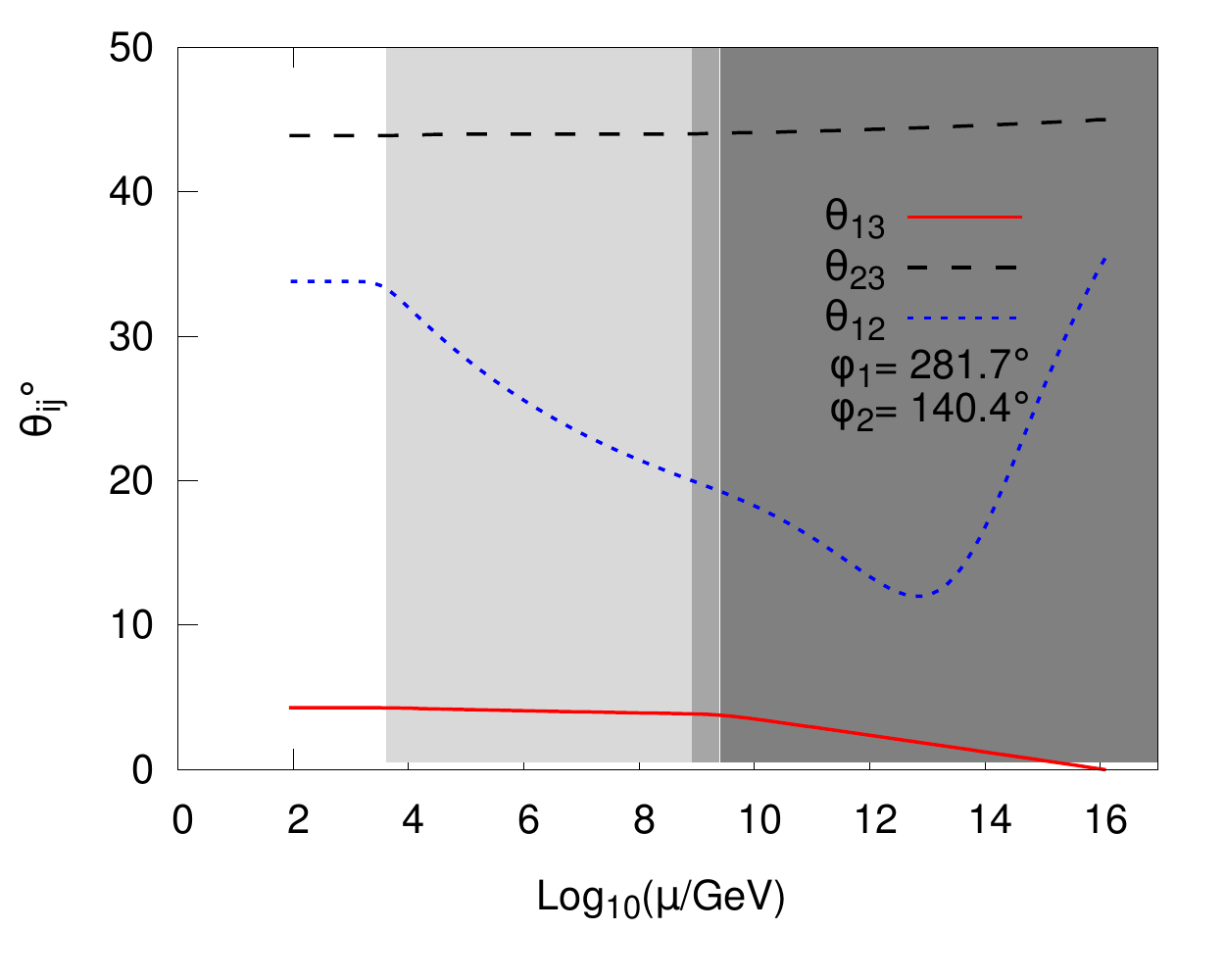}
\includegraphics[width=0.42\textwidth]{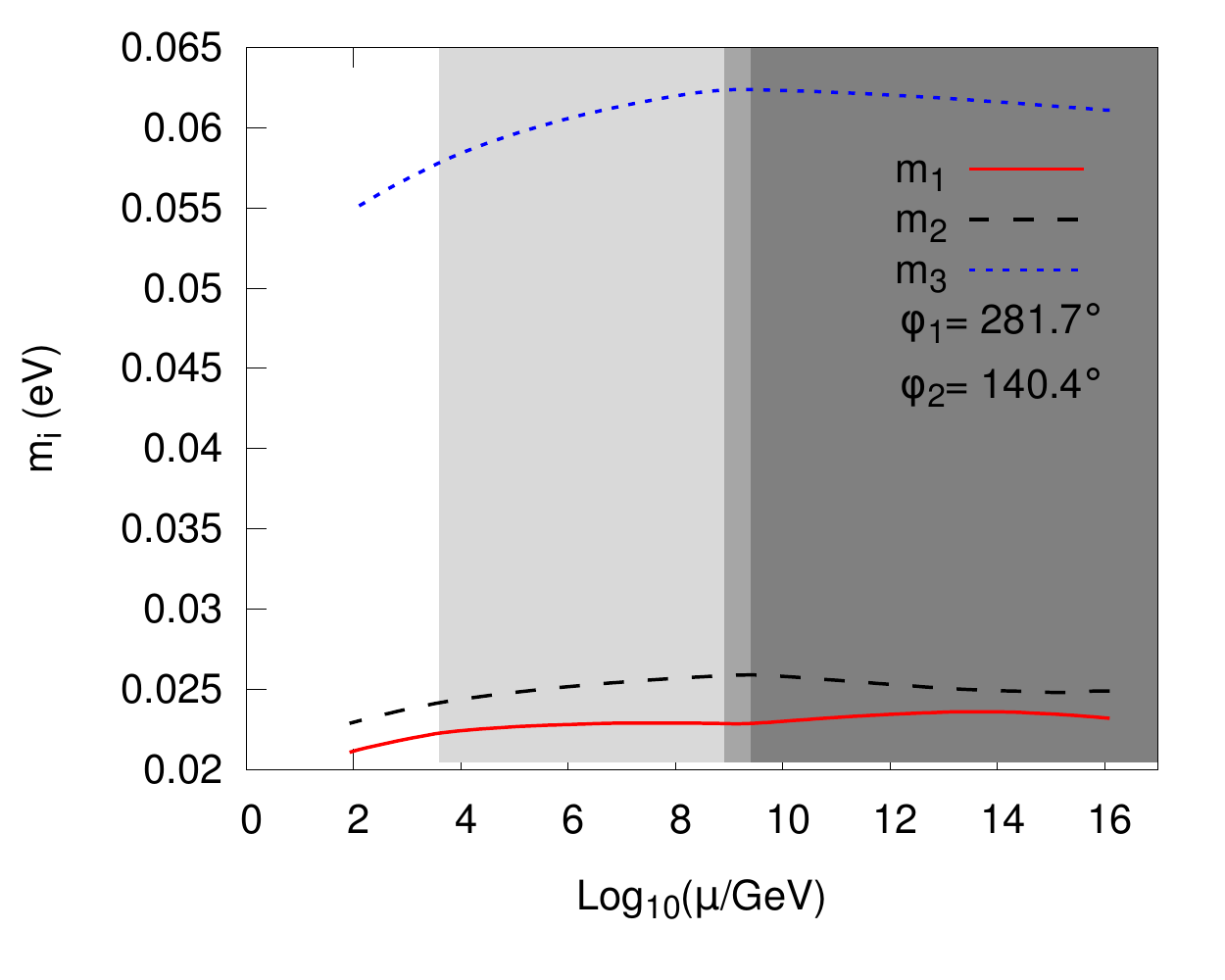}
\end{center}
 \begin{center}
\caption{The RGE of mixing angles and masses in the SM for TBM mixing. The input parameters are given in the second and third column
 of Table 3. The grey shaded areas illustrate the ranges of effective theories when heavy right handed singlets are integrated out.}
 \end{center}
\end{figure}

\subsection{RGE and seesaw threshold corrections in the SM}

\begin{table}[h]
\begin{small}
\label{resultsbm1}
\begin{center}
\begin{tabular}{c|cc|c|cc}
\hline
\hline
SM Input  & $\varphi_1,\varphi_2=0$   &$\varphi_1,\varphi_2\neq0$  & SM Output & $\varphi_1,\varphi_2=0$  & $\varphi_1,\varphi_2\neq0$  \\
\hline
$r_1$ & 0.64$\times 10^{-2} $ & 0.24$\times 10^{-2}$ & $M_{R_1}$ (GeV) & 3$\times$10$^5$ & 6.5$\times$10$^5$\\
$r_2$ & 0.65 & 0.703  & $M_{R_2}$ (GeV) & 1.27$\times$10$^9$ & 5.8 $\times$10$^9$ \\
$\delta$ & 243$^\circ $ & 268$^\circ $  & $M_{R_3}$ (GeV) & 2.4$\times$ 10$^{9} $ & 8$\times$ 10$^{10}$\\
$y_{\nu}$ & 0.37&  0.74 & -& -&-\\
$\theta_1$ & 243$^\circ$  & 163$^\circ$ & $\theta_{12}$ & 34.7$^\circ$ & 33$^\circ$  \\
$\theta_2$ & 60.2$^\circ$ &329$^\circ$ & $\theta_{23}$ & 43$^\circ$ & 50.2$^\circ$ \\
$\theta_3$ & 306$^\circ$ & 333.5$^\circ$  & $\theta_{13}$ & 3.52$^\circ$ & 5.07$^\circ$\\
$ m_1$ (eV) & 0.0264 & 4.8$ \times 10^{-3}$ & $ m_1$ (eV) & 0.0243 &1.58$ \times 10^{-3}$\\
$\Delta m^2_{12}$(eV$^2$) & 1.5$\times$ 10$^{-4}$ &  4.8$\times$ 10$^{-7}$ & $\Delta m^2_{12}$(eV$^2$) & 7.03 $\times$ 10$^{-5}$ & 7.94 $\times$ 10$^{-5}$ \\
$\Delta m^2_{13}$(eV$^2$) & 3.07$\times$ 10$^{-3}$ & 3.4$\times$ 10$^{-3}$ & $\Delta m^2_{13}$(eV$^2$) & 2.4$\times$ 10$^{-3}$  &  2.36$\times$ 10$^{-3}$ \\
$\varphi_1$  & 0$^\circ$ & 7.85$^\circ$ & $\varphi_1$ & 4$^\circ$ &2.8$^\circ$  \\
$\varphi_2$ & 0$^\circ$ & 112.6$^\circ$ & $\varphi_2$ & 3.05$^\circ$ & 90$^\circ$ \\
--&--&--& $J_{CP}$ & -7.3$\times10^{-4}$ & -0.01\\
--&--& --& $M_{ee} (eV)$ & 4.14$\times10^{-3}$ & 0.024 \\
\hline
\end{tabular}
\end{center}
\begin{center}
\caption{
Numerical values of input and output parameters radiatively generated in the SM for BM mixing
for zero and non zero Majorana phases at $\Lambda_{\rm GUT}$ = 2$\times 10^{16}$GeV.
The input values for neutrino mixing angles at GUT scale are $\theta_{13}$=0$^\circ$, $\theta_{23}$= $\theta_{12}= $45$^\circ$.}
 \end{center}
 \end{small}
\end{table}
\begin{figure}[h]
\begin{center}
\label{figbmsm}
\includegraphics[width=0.42\textwidth]{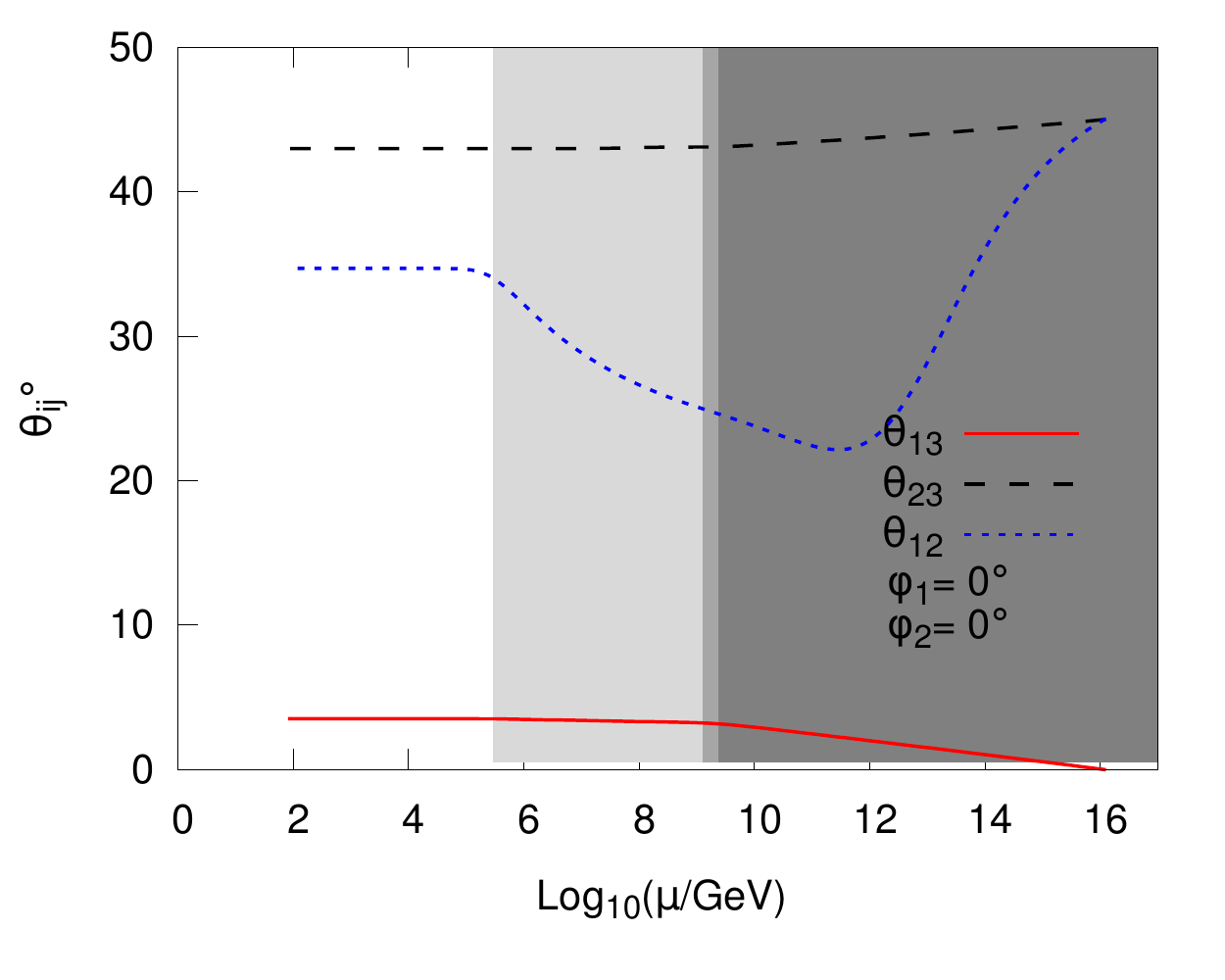}
\includegraphics[width=0.42\textwidth]{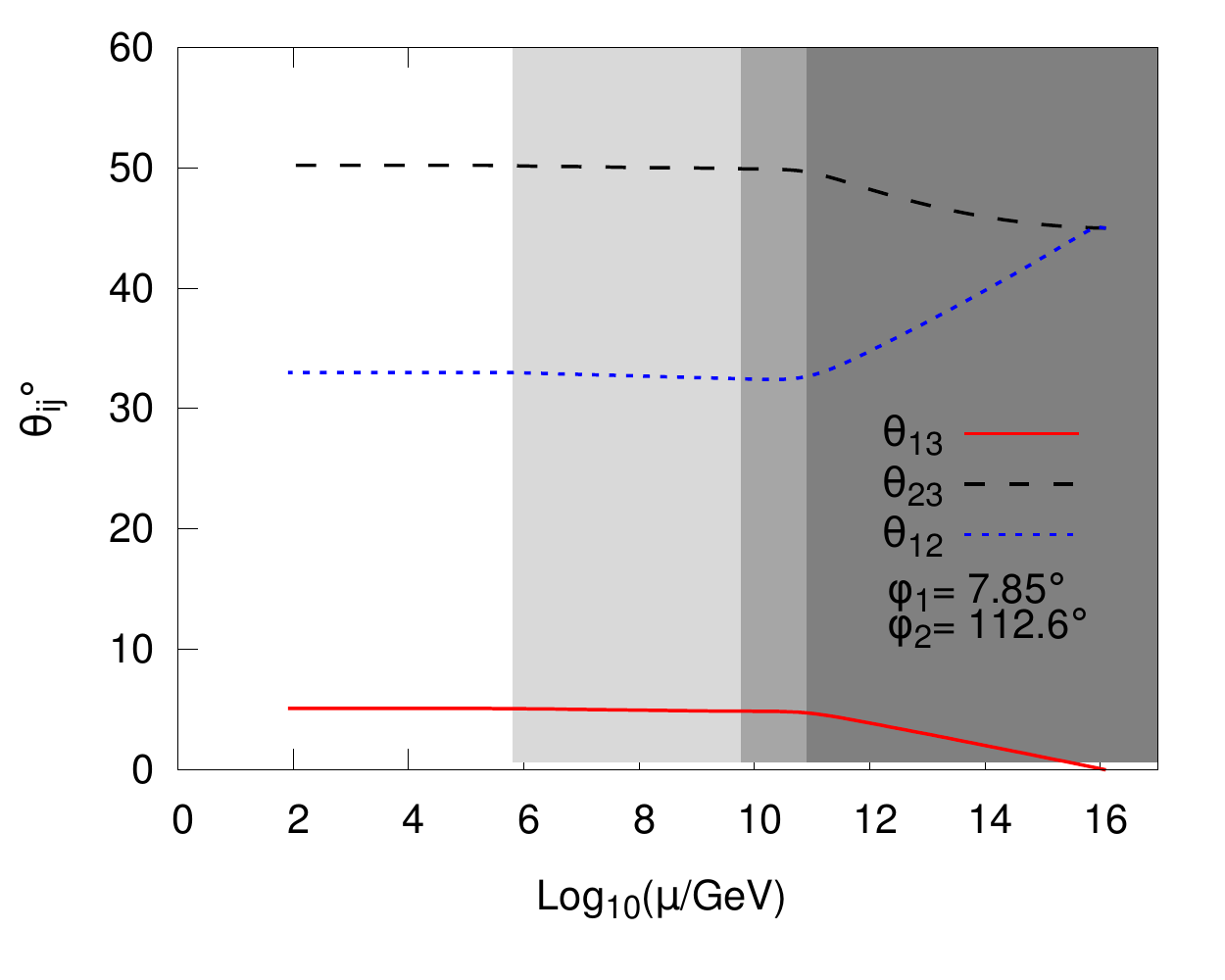}
\end{center}
 \begin{center}
\caption{The RGE of mixing angles in the SM for BM mixing. The input
parameters are given in the second and third column of Table 4. The grey shaded areas
illustrate the ranges of effective theories when heavy right handed singlets are integrated out.}
 \end{center}
\end{figure}
In this analysis we start with different neutrino mass matrices, $M_{\nu}$, that are diagonalized
 by the mixing matrices, $U$ given in Table 2, respectively.
Running of the RGE can be divided into three regions
governed by different RG equations in the respective regions as\\
a) from $\Lambda_{\rm GUT}$ down to the highest seesaw scale $M_{R_3}$, \\
b) in between the three seesaw scales, \\
c) from the lowest seesaw scale $M_{R_1}$ down to $\Lambda_{\rm EW}$. \\
At the leading order the expression for $\theta_{12}$ is
inversely proportional to solar mass squared
difference ($\Delta m_{12}^2$), whereas the other mixing angles $\theta_{23}$ and $\theta_{13}$ are both inversely
related to atmospheric mass squared difference ($\Delta m_{13}^2$).
Thus, $\theta_{12}$ is maximally affected by the RGE
among all the mixing angles. In case of quasidegenerate neutrino spectrum there can be
visible corrections to other mixing angles also. In the SM, RGE corrections to neutrino mixing angles in absence of
threshold corrections are negligible. However, inclusion of the threshold effects can
significantly modify RGE of neutrino masses.
\begin{table} [h]
\begin{small}
\label{resultshm1}
\begin{center}
\begin{tabular}{c|cc|c|cc}
\hline
\hline
SM Input  &  $\varphi_1,\varphi_2=0$  &  $\varphi_1,\varphi_2\neq0$  & SM Output & $\varphi_1,\varphi_2=0$ &   $\varphi_1,\varphi_2\neq0$  \\
\hline
$r_1$ &0.29$\times 10^{-2} $ & 0.63$\times 10^{-2} $ & $M_{R_1}$ (GeV) & 6.4$\times$10$^5$& 4.7$\times$10$^5$ \\
$r_2$ & 0.57 & 0.68 & $M_{R_2}$ (GeV)& 3.3 $\times$10$^9$& 3.1$\times$10$^9$ \\
$\delta$ & 23.1$^\circ $ &337.5$^\circ $ & $M_{R_3}$ (GeV)&3.7 $\times$ 10$^{10} $&7.6 $\times$ 10$^{9}$\\
$y_{\nu}$& 0.661& 0.59 & -& -&-\\
$\theta_1$ &146$^\circ$  & 147.2$^\circ$  &$\theta_{12}$ & 33.7$^\circ$ & 34.4$^\circ$  \\
$\theta_2$ &261$^\circ$ & 271.6$^\circ$ & $\theta_{23}$ & 45.2$^\circ$& 44.8$^\circ$\\
$\theta_3$ & 175.3$^\circ$ & 92.25$^\circ$ &$\theta_{13}$ & 1.4$^\circ$&6.9$^\circ$\\
$ m_1$ (eV)& 4.14$ \times 10^{-3}$& 0.0294 &$ m_1$ (eV) & 3.38$ \times 10^{-3}$ & 0.0245 \\
$\Delta m^2_{12}$(eV$^2$) & 9.6 $\times$ 10$^{-5}$ &  8.6 $\times$ 10$^{-5}$ & $\Delta m^2_{12}$(eV$^2$)& 7.4 $\times$ 10$^{-5}$&7.63 $\times$ 10$^{-5}$ \\
$\Delta m^2_{13}$(eV$^2$)&3.65 $\times$ 10$^{-3}$ & 3.8 $\times$ 10$^{-3}$ & $\Delta m^2_{13}$(eV$^2$) &2.4$\times$ 10$^{-3}$ &2.6$\times$ 10$^{-3}$ \\
$\varphi_1$&0$^\circ$& 340.3$^\circ$  & $\varphi_1$ & 160.2$^\circ$ & 25.9$^\circ$\\
$\varphi_2$&0$^\circ$& 219.6$^\circ$ & $\varphi_2$ & 151.4$^\circ$ & 242$^\circ$\\
--&--&--&$J_{CP}$& -5.36$\times$10$^{-3}$ & -0.024  \\
--&--& --&$M_{ee} (eV)$&4.9$\times$10$^{-3}$ & 0.022  \\
\hline
\end{tabular}
\end{center}
\begin{center}
\caption{Numerical values of input and output parameters radiatively generated in the SM for HM mixing for zero and non zero Majorana phases at $\Lambda_{\rm GUT}$ = 2$\times 10^{16}$GeV. The input values for neutrino mixing angles at GUT scale are $\theta_{13}$= 0$^\circ$, $\theta_{23}$= 45$^\circ$ and $\theta_{12}$= 30$^\circ$.}
 \end{center}
 \end{small}
\end{table}
\begin{figure}[h]
\begin{center}
\label{fighmsm}
\includegraphics[width=0.42\textwidth]{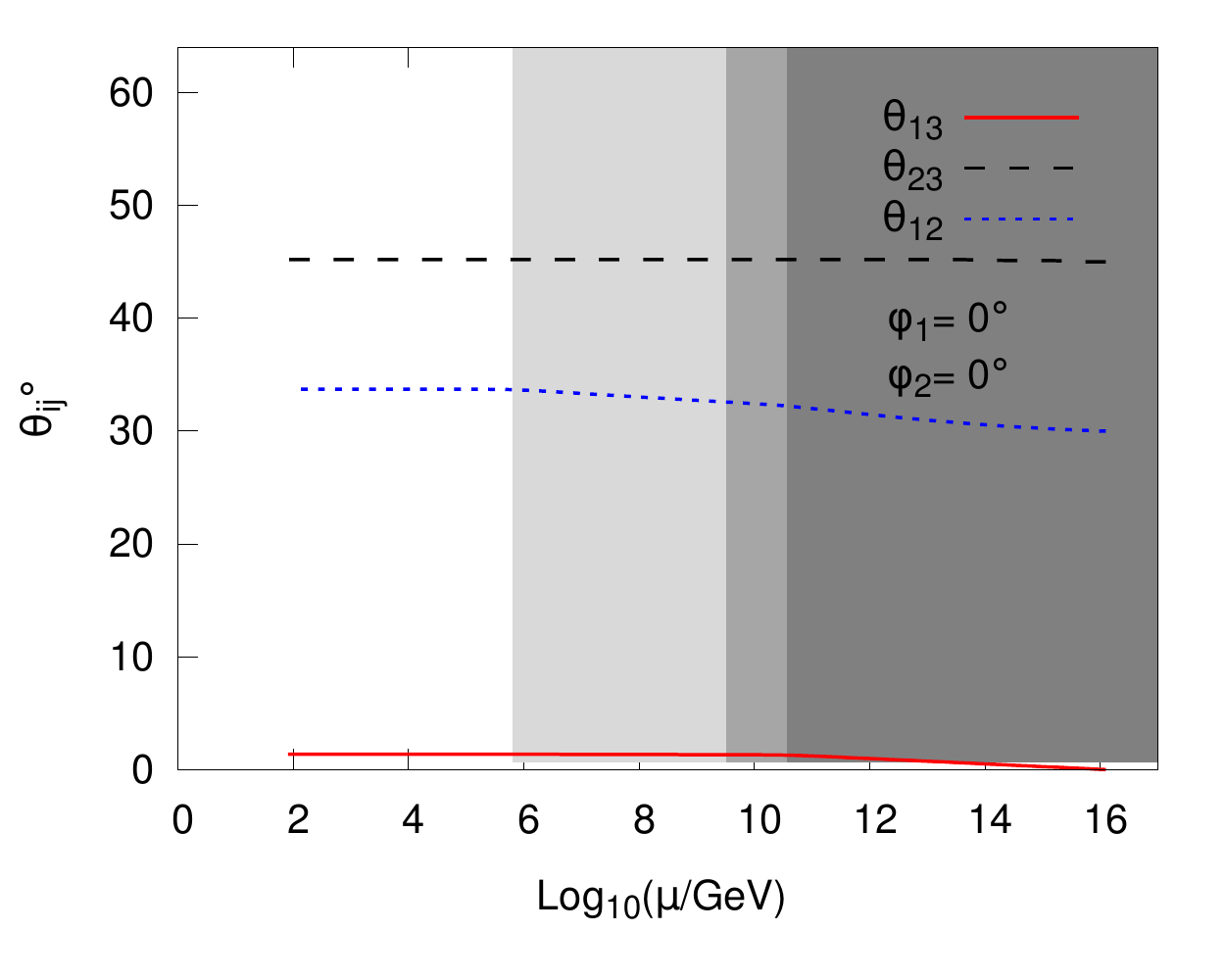}
\includegraphics[width=0.42\textwidth]{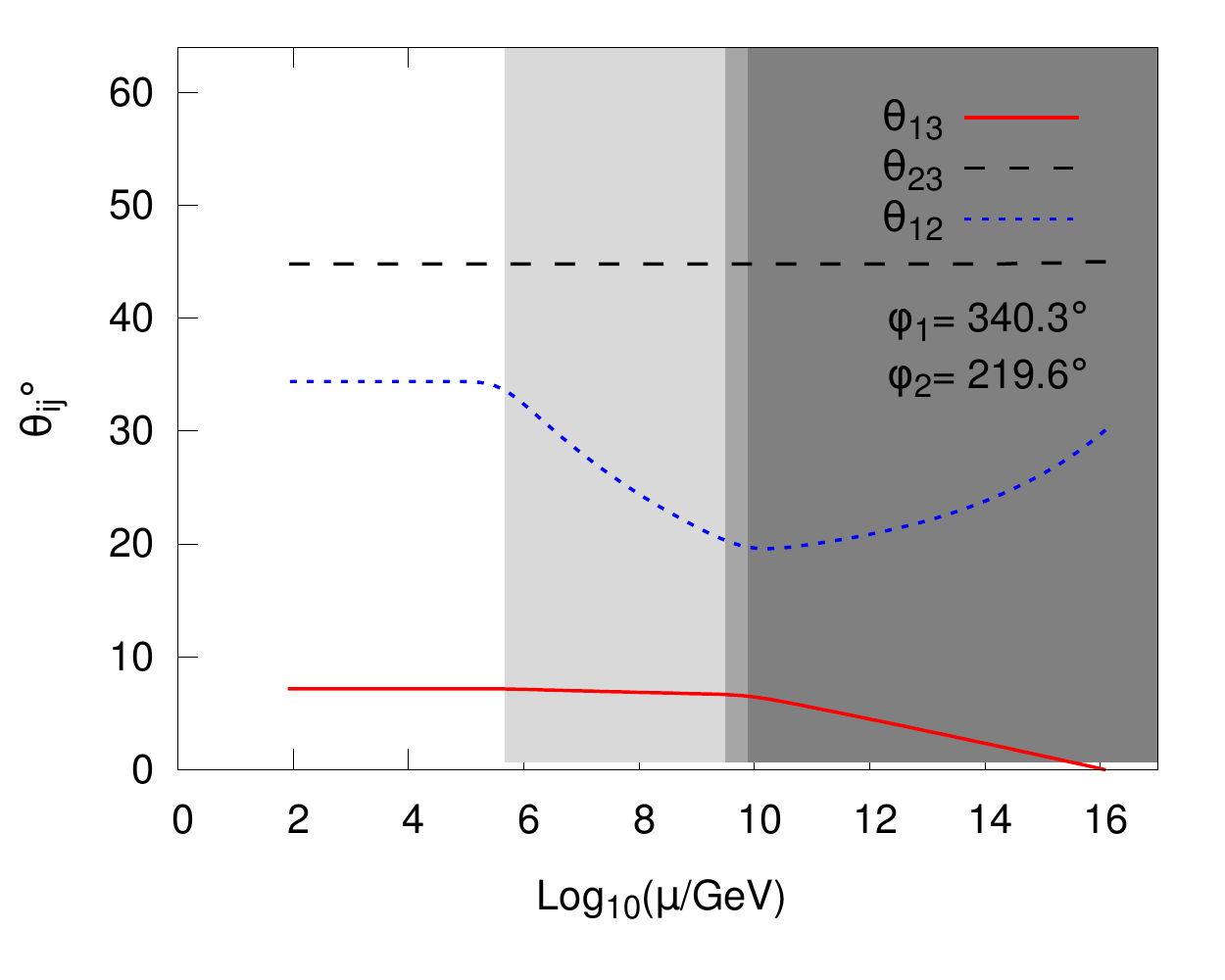}
\end{center}
 \begin{center}
\caption{The RGE of mixing angles in the SM for HM mixing. The input
parameters are given in the second and third column of Table 5. The grey shaded areas illustrate the ranges of effective theories when heavy right handed singlets are integrated out.}
 \end{center}
\end{figure}
We study the RGE of neutrino mixing angles from $\Lambda_{\rm GUT}$ to $\Lambda_{\rm EW}$ in the
SM for zero and non zero Majorana phases in all four mixing scenarios.

The behaviors of RGE for mixing angles $\theta_{12}$, $\theta_{23}$ and $\theta_{13}$ in all cases are shown in Figs. 1, 2, 3 and 4 for TBM, BM, HM and GR, respectively.
In those figures, the left panel corresponds to $\varphi_1=\varphi_2=0$, while the right panel has
non-zero $\varphi_1$ and $\varphi_2$.
The set of input parameters taken at $\Lambda_{\rm GUT}$  corresponding to each cases are given
in the second and third columns of Table 3, 4, 5 and 6, respectively. The last two columns in those tables correspond to the output parameters obtained at $\Lambda_{\rm EW}$.
\begin{table} [h]
\begin{small}
\label{resultsgr1}
\begin{center}
\begin{tabular}{c|cc|c|cc}
\hline
\hline
SM Input  &  $\varphi_1,\varphi_2=0$   & $\varphi_1,\varphi_2\neq0$ & SM Output &  $\varphi_1,\varphi_2=0$  & $\varphi_1,\phi_2\neq0$ \\
\hline
$r_1$ &0.27$\times 10^{-2} $ & 0.68$\times 10^{-3}$ & $M_{R_1}$ (GeV) & 8.7$\times$10$^5$& 4.9$\times$10$^5$\\
$r_2$ & 0.6 & 0.35 &$M_{R_2}$ (GeV)& 4$\times$10$^9$ & 7.7$\times$10$^8$ \\
$\delta$ & 55.2$^\circ $ & 63.0$^\circ$&$M_{R_3}$ (GeV)&6.6$\times$ 10$^{10}$&1.8$\times$ 10$^{9} $\\
$y_{\nu}$& 0.7&  0.65 & -& -&-\\
$\theta_1$ &46.8$^\circ$  & 192.5$^\circ$ &$\theta_{12}$ &34.4$^\circ$ &34.2$^\circ$ \\
$\theta_2$ &225$^\circ$ & 185.6$^\circ$ &$\theta_{23}$& 44.6$^\circ$& 41.9$^\circ$  \\
$\theta_3$ & 123$^\circ$ & 249$^\circ$  &$\theta_{13}$& 1.73$^\circ$& 6$^\circ$\\
$ m_1$ (eV)& 2.1$ \times 10^{-3}$& 0.079 &$ m_1$ (eV)&1.62$ \times 10^{-3}$& 0.067  \\
$\Delta m^2_{12}$(eV$^2$) & 8.9 $\times$ 10$^{-5}$& 9.3 $\times$ 10$^{-5}$ &$\Delta m^2_{12}$(eV$^2$)& 7.25 $\times$ 10$^{-5}$& 7.6 $\times$ 10$^{-5}$\\
$\Delta m^2_{13}$(eV$^2$)&3.7 $\times$ 10$^{-3}$& 3.8 $\times$ 10$^{-3}$ &$\Delta m^2_{13}$(eV$^2$) &2.3$\times$ 10$^{-3}$ &2.3$\times$ 10$^{-3}$  \\
$\varphi_1$&0$^\circ$& 240.6$^\circ$&$\varphi_1$&171$^\circ$ &13.6$^\circ$\\
$\varphi_2$&0$^\circ$& 353.5$^\circ$&$\varphi_2$&185$^\circ$ & 107.5$^\circ$\\
--&--&--&$J_{CP}$& 0.007 & -0.02\\
--&--& --&$M_{ee} (eV)$&3.8$\times$10$^{-3}$  & 0.06\\
\hline
\end{tabular}
\end{center}
\begin{center}
\caption{Numerical values of input and output parameters radiatively generated in the SM for GR mixing
for zero and non zero Majorana phases at $\Lambda_{\rm GUT}$ = 2$\times 10^{16}$GeV.
The input neutrino mixing angles at GUT scale are $\theta_{13}$= 0$^\circ$, $\theta_{23}$= 45$^\circ$ and $\theta_{12}$= 31.7$^\circ$.}
 \end{center}
 \end{small}
\end{table}
\begin{figure}[h]
\begin{center}
\label{figgrsm}
\includegraphics[width=0.42\textwidth]{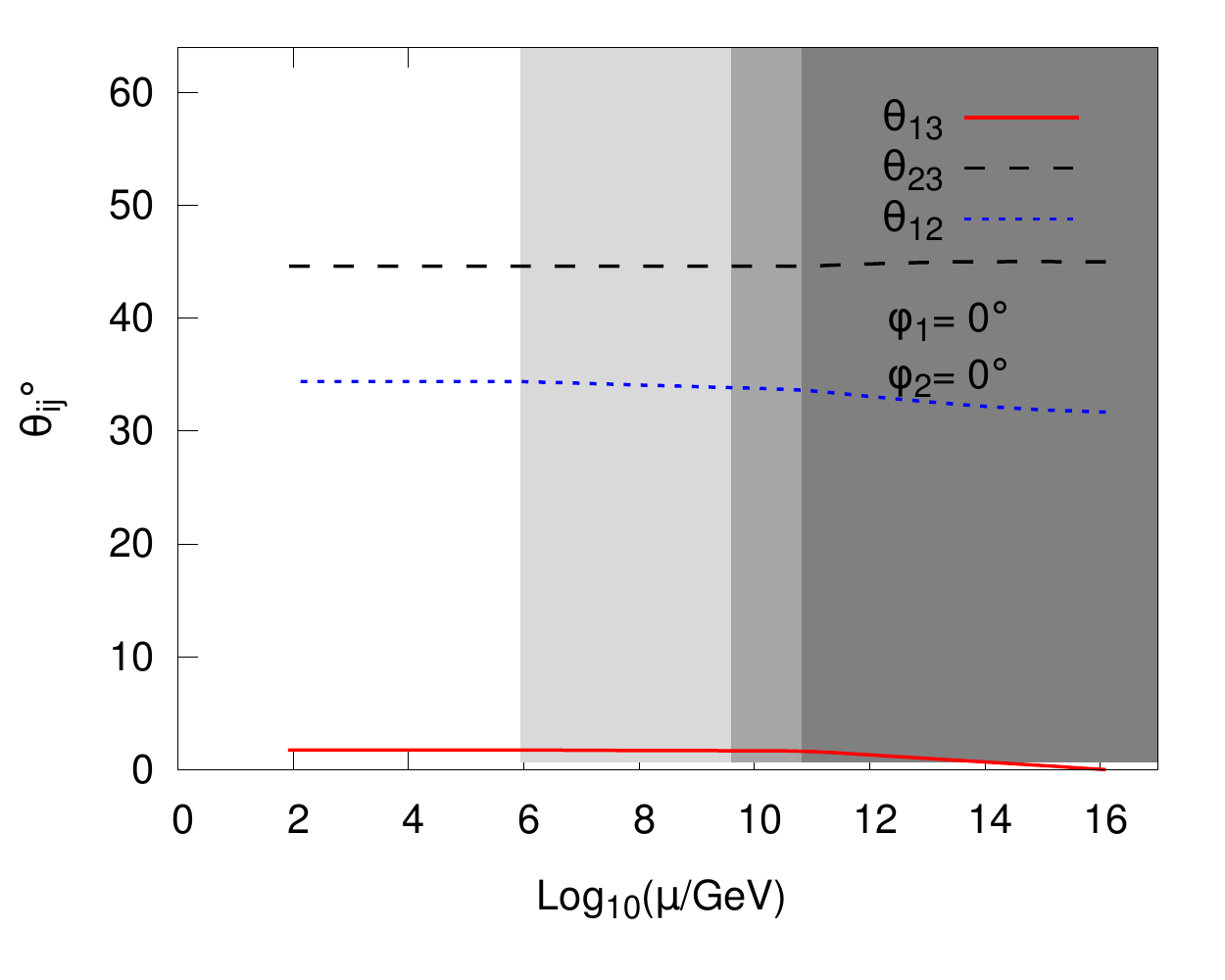}
\includegraphics[width=0.42\textwidth]{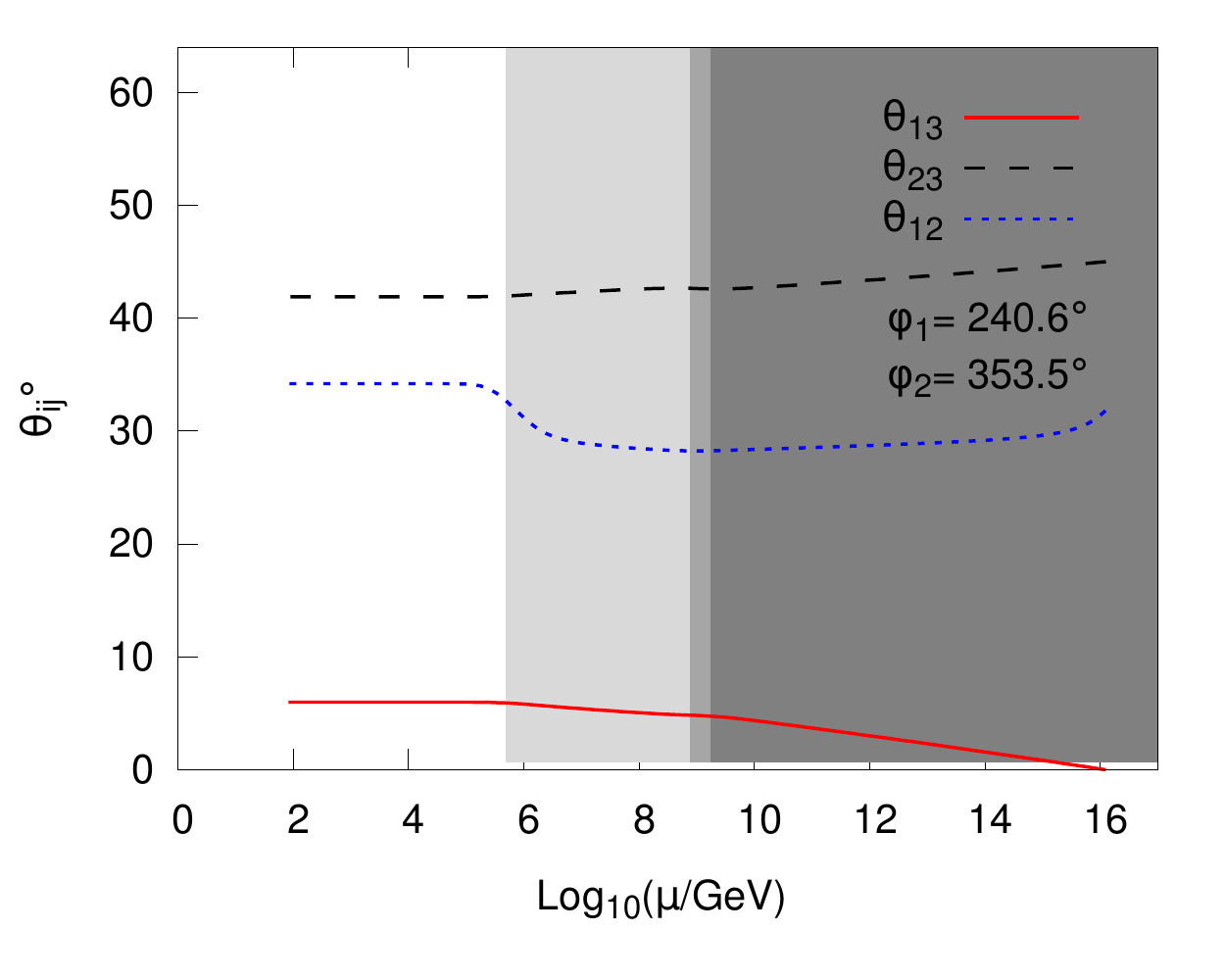}
\end{center}
 \begin{center}
\caption{The RGE of mixing angles in the SM for GR mixing. The input
parameters are given in the second and third column of Table 6. The grey shaded areas illustrate the ranges of
 effective theories when heavy right handed singlets are integrated out.}
 \end{center}
\end{figure}
The effects of RGEs for the neutrino mixing angles below the lowest seesaw scale  $M_{R_1}$ down to $\Lambda_{\rm EW}$ are negligible in the SM because of small corrections arisen only due to $Y_l$.

However, at the energy scale between and above the seesaw scales, there will be additional contributions of $Y_{\nu}$
along with $Y_{l}$. Thus, the RGE  is dependent on $Y_{\nu}$ which is free and can be as large as $\mathcal{O}(1)$.
Heavy right handed fields are subsequently integrated out at the three seesaw scales shown by the three grey regions in the figures.
Thus (n-1)$\times$3 submatrix of $Y_{\nu}$ remains after each step of integrating out M$_{R_i}$.
As can be seen from Eq.(\ref{mnu3}), the running between the seesaw scales is dependent on the sum of two terms
$\kappa^{(n)}$ and $2Y_{\nu}^{T(n)}M_{R}^{(n)}Y_{\nu}^{(n)}$. As discussed in \cite{Antusch:2005gp},
in the SM the RGE scaling in these two terms
is different due to interaction with trivial flavor structure. This implies that there
can be large corrections for the mixing angles between these threshold scales in the SM.
The values of three seesaw scales are given in the output column of Tables 3, 4, 5 and 6.
From Eq.({\ref{Mr1}}), the heavy right handed Majorana masses in the TBM at the GUT scale for vanishing Majorana phases are found to be $M_{R_i}$= 2.62$\times$10$^4$eV, 2.23$\times$10$^{9}$eV, 1.75$\times$10$^{11}$eV respectively. We observe small running of these values between GUT and seesaw scales.
There are significant corrections to mixing angles especially $\theta_{12}$ between
and above the seesaw scales for non zero Majorana phases in the SM, whereas small running for vanishing $\varphi_1$ and $\varphi_2$ as shown in Figs. 1, 2, 3 and 4. We found that small value of $\theta_{13} < $3.5$^\circ$ is obtained for
 vanishing Majorana phases. However, when these phases are non zero $\theta_{13}$ as large as 6.9$^\circ$ is
 produced. In all four cases, there is considerable corrections to $\theta_{12}$
 and it is possible to obtain it's value near the best fit (33$^\circ$)
at the low scale when it ranges from (30$^\circ$- 45$^\circ$) at the high GUT scale.
In the absence of the Majorana phases it is not possible to have large values of $\theta_{13}$ at the low scale in the SM.
In Ref. \cite{xing} it has been shown that in the SM $\theta_{13}$ as large as 5$^\circ$ can only be obtained when very large $\theta_{12}$=67$^\circ$ is considered at the GUT scale. However, taking into consideration the Majorana
phases, $\theta_{13}$ as large as $\approx$ 5$^\circ$-- 6.9$^\circ$ is obtained at the low scale when $\theta_{12}$ at $\Lambda_{\rm GUT}$ is in the range of (30$^\circ$- 45$^\circ$).
Thus, the Majorana phases can significantly affect the RGE  of neutrino
mixing angles \cite{haba} as observed above. However, in the SM, the value of $\theta_{13}$ is still below 3$\sigma$ allowed range at low scale for both zero and non zero Majorana phases.

In Fig. 1 we also show the RGE  of the neutrino masses from $\Lambda_{\rm GUT}$ to $\Lambda_{\rm EW}$ for both
zero and non zero Majorana phases in the TBM. The running of the mass eigenvalues in this region below the seesaw
scales can be significant in the SM due to
the factor $\alpha$ (Eq.\ref{alp}) which can be larger than $Y_{\tau}^2$. As we see from the right panel of Fig.1 there is
running of masses even below $M_{R_1}$, irrespective of values of $\varphi_1$ and $\varphi_2$. It indicates running of masses is not directly dependent on the Majorana phases \cite{casas2}.
Due to radiative generation of non zero values of $\theta_{13}$ and $\delta_{CP}$ below the GUT scale, non vanishing values
of Jarlskog rephasing invariant are generated at $\Lambda_{\rm EW}$, which lead to observable CP
violation in neutrino oscillation experiments.
For the best fit values of mixing angles and Dirac phase $\delta_{CP}$ (300$^\circ$) given in the global analysis \cite{garcia}, the Jarlskog Invariant is determined to be J$_{CP}= -0.028$. In the SM for non vanishing Majorana phases we obtain J$_{CP}$ $\sim$ -10$^{-2}$ at the EW scale for BM, HM and GR scenarios. The measurement of $\delta_{CP}$ from long baseline neutrino oscillation experiments in future would be useful to study the viability of these mixing scenarios at high scale.

Neutrinoless double beta decay (0$\nu \beta \beta$) if observed, would imply lepton number violation (LNV) and Majorana nature of neutrinos. The current experimental results for 0$\nu \beta \beta$ can constrain the effective Majorana neutrino mass, $M_{ee}$.
From the search for 0$\nu \beta \beta$ of $^{136}$Xe at EXO-200 \cite{exo},
the effective Majorana mass $M_{ee}$ is constrained to be less than (0.14--0.38)eV at 90\% C.L.. A combination of limits from KamLAND-Zen \cite{Gando} and EXO-200 constrains this limit further to less than (0.12--0.25)eV at 90\% C.L. based on representative range of available matrix element calculations. The predictions of $M_{ee}$ for all for four mixing scenarios are given in corresponding tables.
 Here $M_{ee}\sim$10$^{-3}$eV is obtained for vanishing phases while $\sim$10$^{-2}$eV is obtained when Majorana phases contribute to the RGE. The observations of the signal in the present and future 0$\nu \beta \beta$ experiments will be crucial to decide the fate of these scenarios under consideration.
 \begin{table}[h]
 \begin{small}
\label{resultstbm1}
\begin{center}
\begin{tabular}{c|cc|c|cc}
\hline
\hline
MSSM Input  & $\varphi_1,\varphi_2=0$  &  $\varphi_1,\varphi_2\neq0$  & MSSM Output & $\varphi_1,\varphi_2=0$  &   $\varphi_1,\varphi_2\neq0$  \\
\hline
$r_1$ & 0.36$\times 10^{-3}$ & 0.42$\times 10^{-3}$ & $M_{R_1}$ (GeV) & 9.9$\times$10$^3$ & 9.13$\times$10$^3$ \\
$r_2$ &  0.47 & 0.68 & $M_{R_2}$ (GeV)&  2.1 $\times$10$^9$ & 2.04 $\times$10$^9$\\
$\delta$ & 238$^\circ$ & 196$^\circ$ & $M_{R_3}$ (GeV)&4.0 $\times$ 10$^{10}$ & 1.36 $\times$ 10$^{10}$\\
$y_{\nu}$&  0.56&  0.46 & -& -&-\\
$\theta_1$ & 176$^\circ$ & 300.2$^\circ$ & $\theta_{12}$ &35.2$^\circ$ & 34.3$^\circ$ \\
$\theta_2$ & 256$^\circ$ & 13.06$^\circ$  &$\theta_{23}$ &49.5$^\circ$ & 40.6$^\circ$  \\
$\theta_3$ & 66.5$^\circ$ & 124.9$^\circ$   &$\theta_{13}$&3.46$^\circ$& 9.46$^\circ$\\
$ m_1$ (eV)&  3.4$\times 10^{-3}$ & 4.5$ \times 10^{-3}$ &$ m_1$ (eV)&2.2$\times 10^{-3}$ & 5.9$\times 10^{-3} $\\
$\Delta m^2_{12}$(eV$^2$) &  2.1 $\times$ 10$^{-5}$& 7.33 $\times$ 10$^{-5}$  &$\Delta m^2_{12}$(eV$^2$)&8 $\times$ 10$^{-5}$&7.48 $\times$ 10$^{-5}$\\
$\Delta m^2_{13}$(eV$^2$)& 2.5 $\times$ 10$^{-3} $ & 3.56$\times 10^{-3}$ &$\Delta m^2_{13}$(eV$^2$) &2.56$\times$ 10$^{-3}$ &2.57$\times$ 10$^{-3}$\\
$\varphi_1$ & 0$^\circ$ & 256.8$^\circ$ &$\varphi_1$& 50.0$^\circ$& 112.7$^\circ$\\
$\varphi_2$&0$^\circ$ & 210.8$^\circ$ & $\varphi_2$ & 30$^\circ$ & 10.5$^\circ$\\
--&--&--&$J_{CP}$ & -3.6 $\times$10$^{-3}$ & -0.0156\\
--&--& --&$M_{ee} (eV)$& 3.3$\times$10$^{-3}$ & 3.7$\times$10$^{-3}$ \\
\hline
\end{tabular}
\end{center}
\begin{center}
\caption{Numerical values of input and output parameters radiatively generated in the MSSM for TBM mixing for zero and nonzero Majorana phases at $\Lambda_{\rm GUT}$ = 2$\times 10^{16}$GeV and tan$\beta$=10. The input values for neutrino mixing angles at GUT scale are $\theta_{13}$= 0$^\circ$, $\theta_{23}$= 45$^\circ$ and $\theta_{12}$= 35.3$^\circ$.}
 \end{center}
 \end{small}
\end{table}
\begin{figure}[h]
\begin{center}
\label{figtbmmssm}
\includegraphics[width=0.42\textwidth]{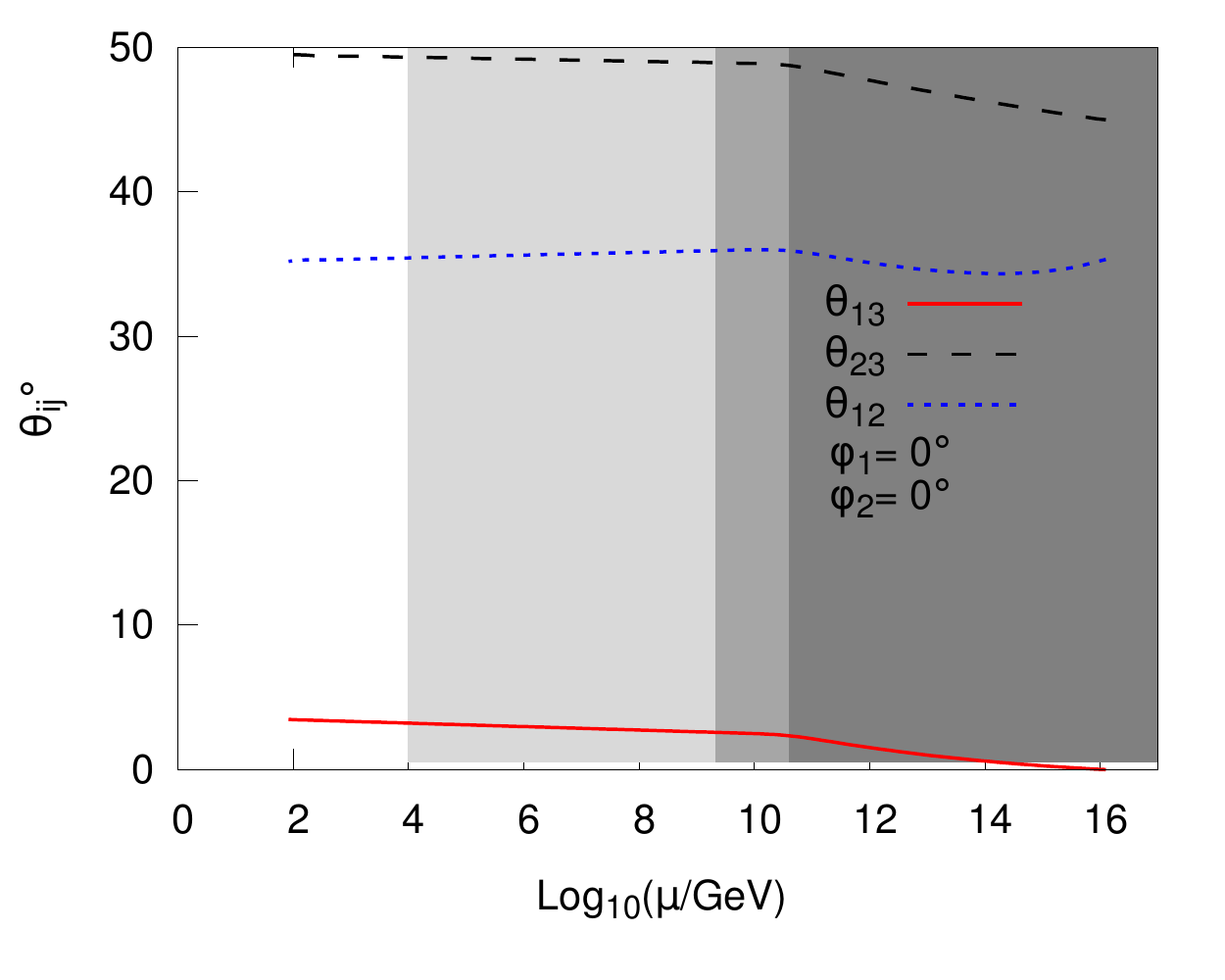}
\includegraphics[width=0.42\textwidth]{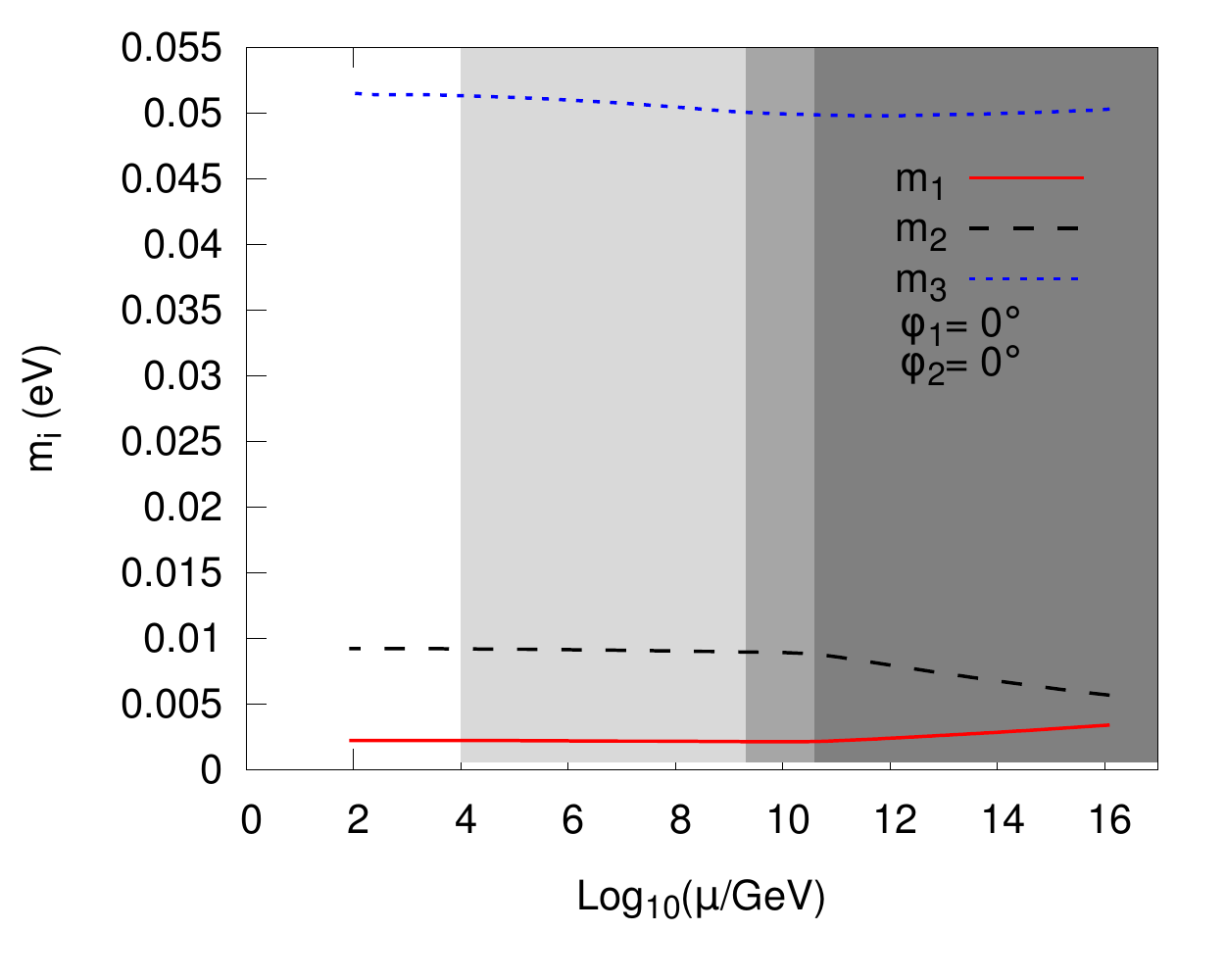}
\includegraphics[width=0.42\textwidth]{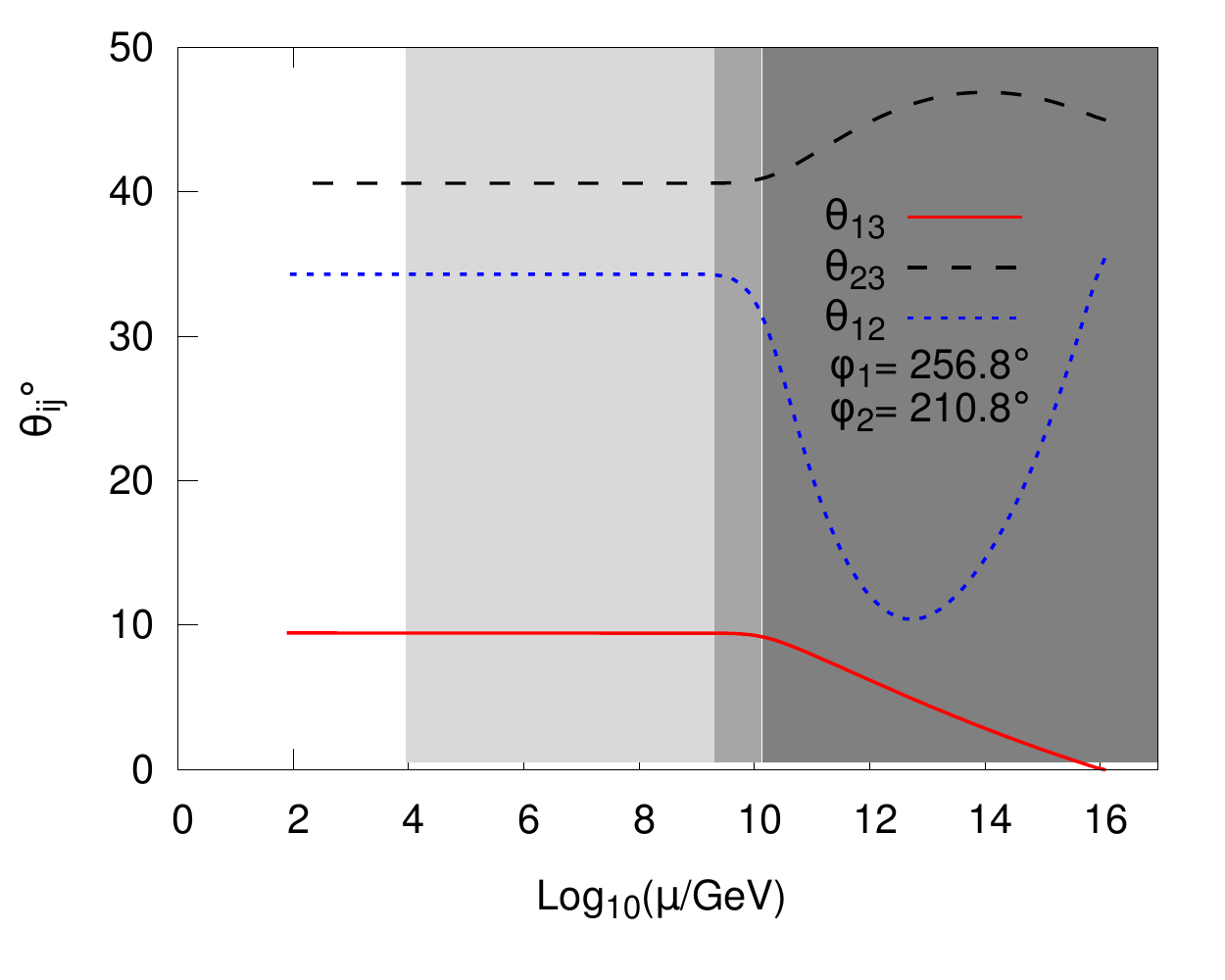}
\includegraphics[width=0.42\textwidth]{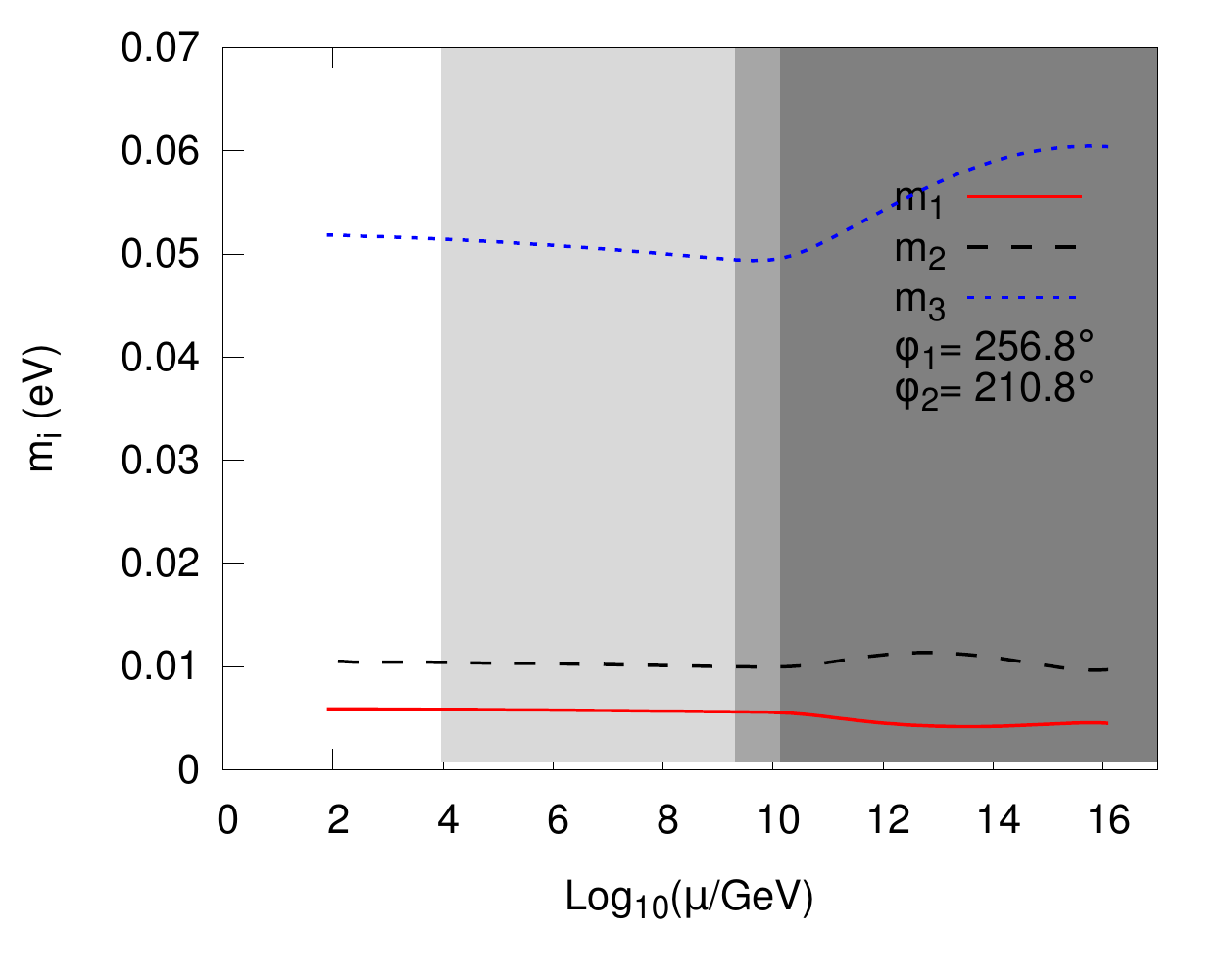}
\end{center}
 \begin{center}
\caption{The RGE  of the mixing angles and masses in the MSSM with tan$\beta$=10. The input parameters are given in the second and third column of Table 7.
The grey shaded areas illustrate the ranges of effective theories when heavy right handed singlets are integrated out.}
 \end{center}
\end{figure}
\begin{table}[h]
\begin{small}
\label{resultsbm1}
\begin{center}
\begin{tabular}{c|cc|c|cc}
\hline
\hline
MSSM Input  & $\varphi_1,\varphi_2=0$  & $\varphi_1,\varphi_2\neq0$  & MSSM Output & $\varphi_1,\varphi_2=0$  & $\varphi_1,\varphi_2\neq0$  \\
\hline
$r_1$ & 0.274$\times 10^{-3}$ & 0.59$\times 10^{-3}$ & $M_{R_1}$ (GeV) &1.4$\times$10$^4$ & 4.2$\times$10$^4$\\
$r_2$ &  0.59 & 0.54 & $M_{R_2}$ (GeV)  & 2.9$\times$10$^9$& 3.0$\times$10$^9$\\
$\delta$ & 261.3$^\circ$ & 157.5$^\circ$ & $M_{R_3}$ (GeV) & 1.17 $\times$ 10$^{11}$ & 3.2$\times$ 10$^{10}$\\
$y_{\nu}$ & 0.584 & 0.66 & -& -&-\\
$\theta_1$ & 44.3$^\circ$ & 115$^\circ$  & $\theta_{12}$ & 31.5$^\circ$ & 33.4$^\circ$  \\
$\theta_2$ & 352$^\circ$ & 356$^\circ$ & $\theta_{23}$ & 38.7$^\circ$ & 38.8$^\circ$ \\
$\theta_3$ &  68.2$^\circ$  & 296$^\circ$ & $\theta_{13}$ & 3.5$^\circ$& 7.41$^\circ$\\
$ m_1$ (eV) & 1.76$ \times 10^{-3}$ & 5.57$ \times 10^{-3}$ & $ m_1$ (eV)  & 4.6$\times 10^{-4} $& 4.23$\times 10^{-3} $\\
$\Delta m^2_{12}$(eV$^2$) &  5.2 $\times$ 10$^{-7}$ & 4.3$\times$ 10$^{-7}$ & $\Delta m^2_{12}$(eV$^2$)  & 7.95 $\times$ 10$^{-5}$& 7.38 $\times$ 10$^{-5}$\\
$\Delta m^2_{13}$(eV$^2$) &  3.2$\times 10^{-3}$ & 3.45$\times 10^{-3}$ &$\Delta m^2_{13}$(eV$^2$)  & 2.64$\times$ 10$^{-3}$ & 2.28$\times$ 10$^{-3}$\\
$\varphi_1$  &  0$^\circ$ & 253.2$^\circ$ & $\varphi_1$  & 315$^\circ$& 230.2$^\circ$\\
$\varphi_2$ & 0$^\circ$ & 295.3$^\circ$ & $\varphi_2$ & 295$^\circ$ & 235.9$^\circ$\\
--&--&--& $J_{CP}$ &  -0.56$\times10^{-2}$  & -0.0233 \\
--&--& --& $M_{ee} (eV)$ & 2.6$\times$10$^{-3}$ & 4.3$\times$10$^{-3}$ \\
\hline
\end{tabular}
\end{center}
\begin{center}
\caption{Numerical values of input and output parameters radiatively generated in the MSSM for BM mixing for zero and non zero Majorana phases at $\Lambda_{GUT}$ = 2$\times 10^{16}$GeV and tan$\beta$=10. The input values for neutrino mixing angles at GUT scale are $\theta_{13}$=0$^\circ$, $\theta_{23}$= $\theta_{12}= $45$^\circ$.}
 \end{center}
 \end{small}
\end{table}
\begin{figure}[h]
\begin{center}
\label{figbmmssm}
\includegraphics[width=0.42\textwidth]{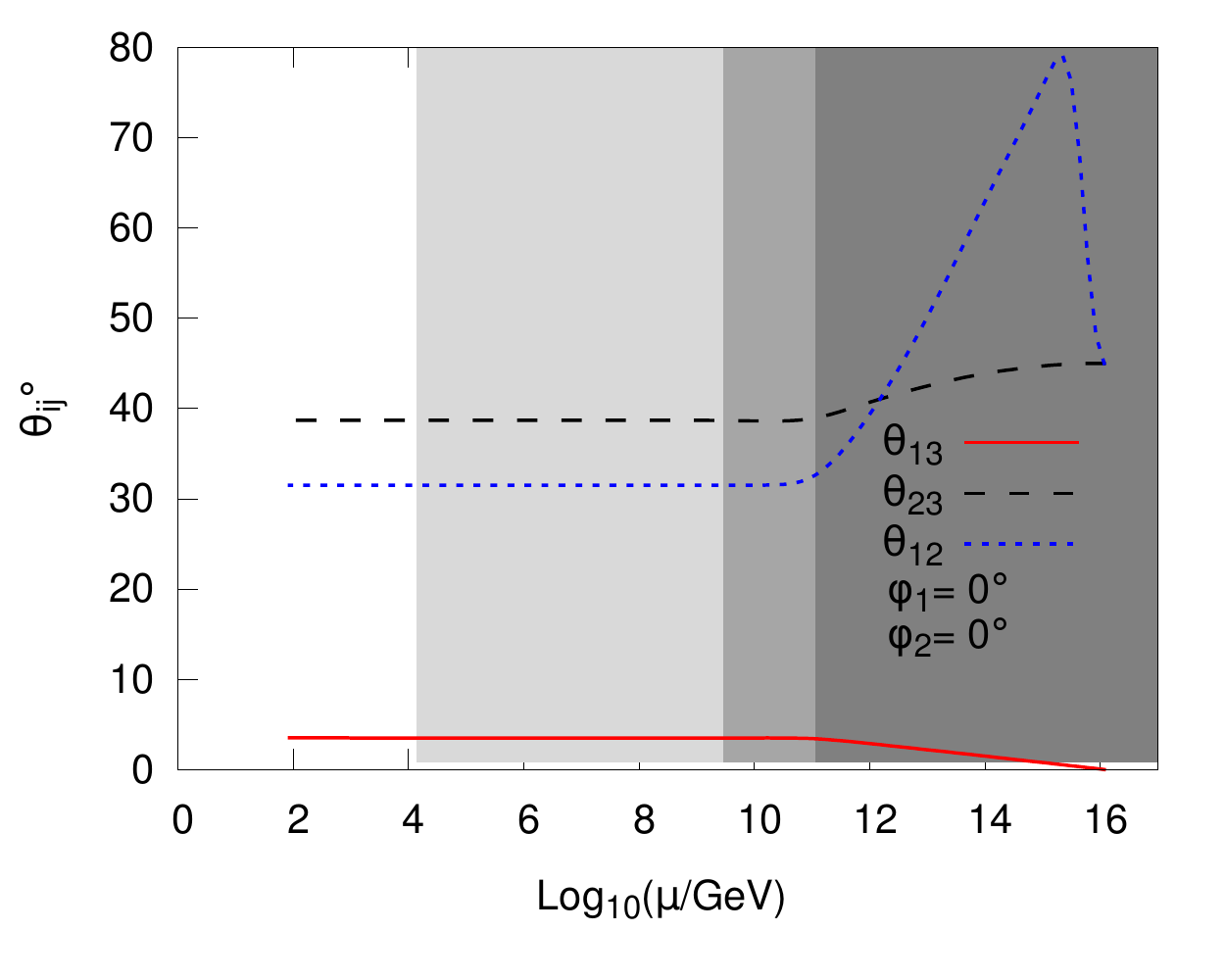}
\includegraphics[width=0.42\textwidth]{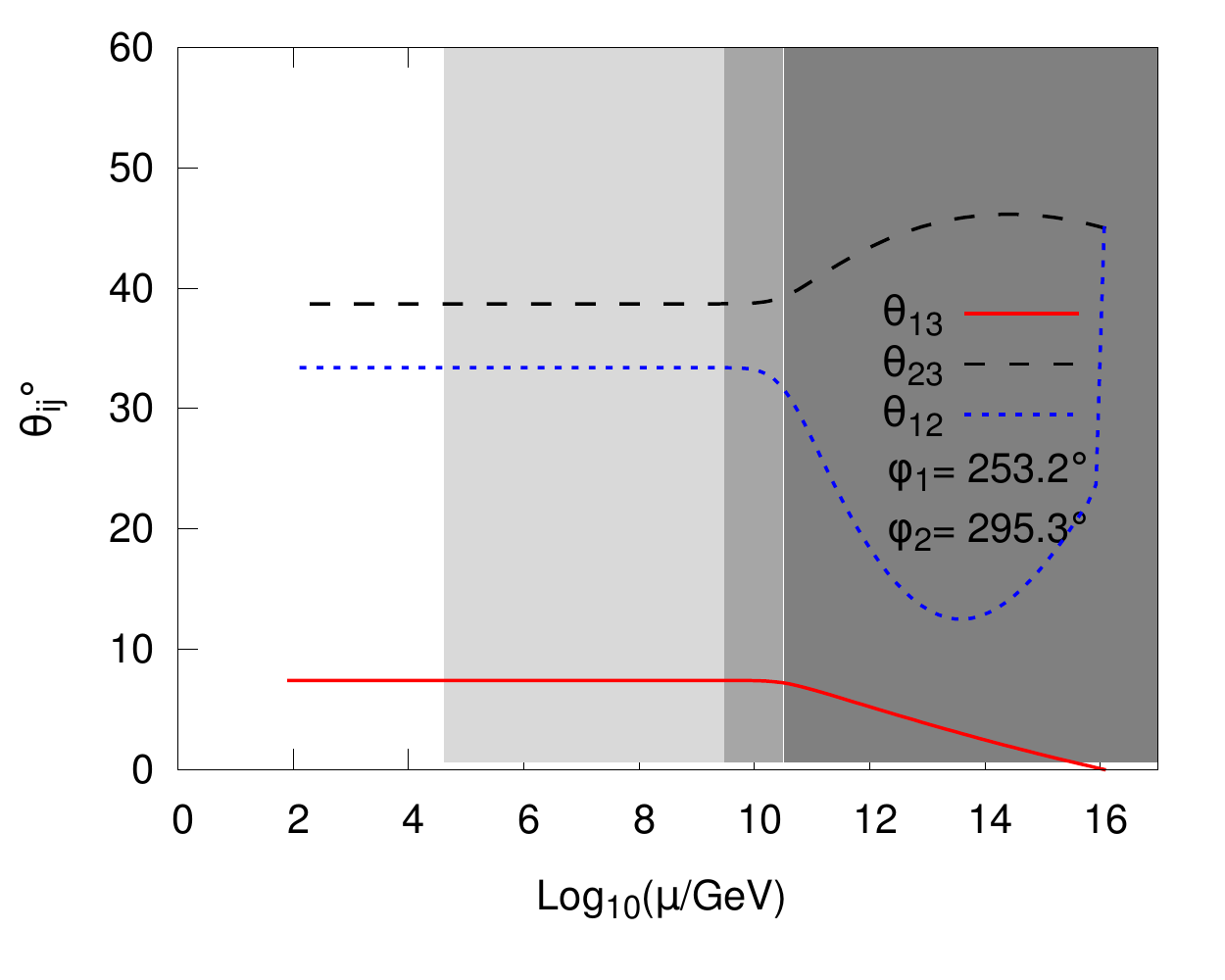}
\end{center}
 \begin{center}
\caption{The RGE  of the mixing angles between $\Lambda_{\rm GUT}$ and $\Lambda_{\rm EW}$ in the MSSM with tan$\beta$=10 for BM mixing.
 The input parameters are given in second and third column of Table 8.
The grey shaded areas illustrate the ranges of effective theories when heavy right handed singlets are integrated out.}
 \end{center}
\end{figure}
\begin{table}[h]
\begin{small}
\label{resultshm2}
\begin{center}
\begin{tabular}{c|cc|c|cc}
\hline
\hline
MSSM Input  &  $\varphi_1,\varphi_2=0$  & $\varphi_1,\varphi_2\neq0$  & MSSM Output & $\varphi_1,\varphi_2=0$  & $\varphi_1,\varphi_2\neq0$  \\
\hline
$r_1$ &0.51$\times 10^{-3} $ & 0.68$\times 10^{-3}$& $M_{R_1}$ (GeV) & 4.36$\times$10$^4$& 5.2$\times$10$^4$ \\
$r_2$ & 0.49 & 0.46&$M_{R_2}$ (GeV)& 2.2$\times$10$^9$& 2.6 $\times$10$^9$\\
$\delta$ & 34.7$^\circ $ & 214.3$^\circ$&$M_{R_3}$ (GeV)&9.5 $\times$ 10$^{10} $&3.3 $\times$ 10$^{10}$\\
$y_{\nu}$ & 0.57&  0.675& -& -&-\\
$\theta_1$ & 126$^\circ$  & 55.6$^\circ$ & $\theta_{12}$ & 33.8$^\circ$ & 35.5$^\circ$  \\
$\theta_2$ & 276$^\circ$ & 145$^\circ$ &$\theta_{23}$ & 39.7$^\circ$& 39.9$^\circ$ \\
$\theta_3$ & 319$^\circ$ & 278.4$^\circ$  & $\theta_{13}$ & 3.7$^\circ$&7.3$^\circ$\\
$ m_1$ (eV) & 2.03$ \times 10^{-3}$ & 5.5$ \times 10^{-3}$ & $ m_1$ (eV) & 5.84$ \times 10^{-4}$ & 3.95$\times 10^{-3} $\\
$\Delta m^2_{12}$(eV$^2$) & 4.8 $\times$ 10$^{-8}$ & 4.7$\times$ 10$^{-8}$  &$\Delta m^2_{12}$(eV$^2$) & 7.45 $\times$ 10$^{-5}$ & 7.6 $\times$ 10$^{-5}$\\
$\Delta m^2_{13}$(eV$^2$)&3.0 $\times$ 10$^{-3}$ & 3.25$\times 10^{-3}$&$\Delta m^2_{13}$(eV$^2$) & 2.6$\times$ 10$^{-3}$ & 2.5$\times$ 10$^{-3}$ \\
$\varphi_1$ & 0$^\circ$ & 304.6$^\circ$ & $\varphi_1$ & 71$^\circ$ & 321.4$^\circ$\\
$\varphi_2$ & 0$^\circ$ & 308.3$^\circ$ & $\varphi_2$ & 99$^\circ$ & 322.2$^\circ$\\
--&--&--&$J_{CP}$& 0.9$\times$10$^{-2}$  & -0.01 \\
--&--& --&$M_{ee} (eV)$&2.5$\times$10$^{-3}$  & 5.2$\times$10$^{-3}$ \\
\hline
\end{tabular}
\end{center}
\begin{center}
\caption{Numerical values of input and output parameters radiatively generated in the MSSM for HM mixing
for zero and non zero Majorana phases at $\Lambda_{GUT}$ = 2$\times 10^{16}$GeV and tan$\beta$=10. The input values for neutrino mixing angles at GUT scale are $\theta_{13}$= 0$^\circ$, $\theta_{23}$= 45$^\circ$ and $\theta_{12}$=30$^\circ$.}
 \end{center}
 \end{small}
\end{table}
\begin{figure}[h]
\begin{center}
\label{fighmmssm}
\includegraphics[width=0.42\textwidth]{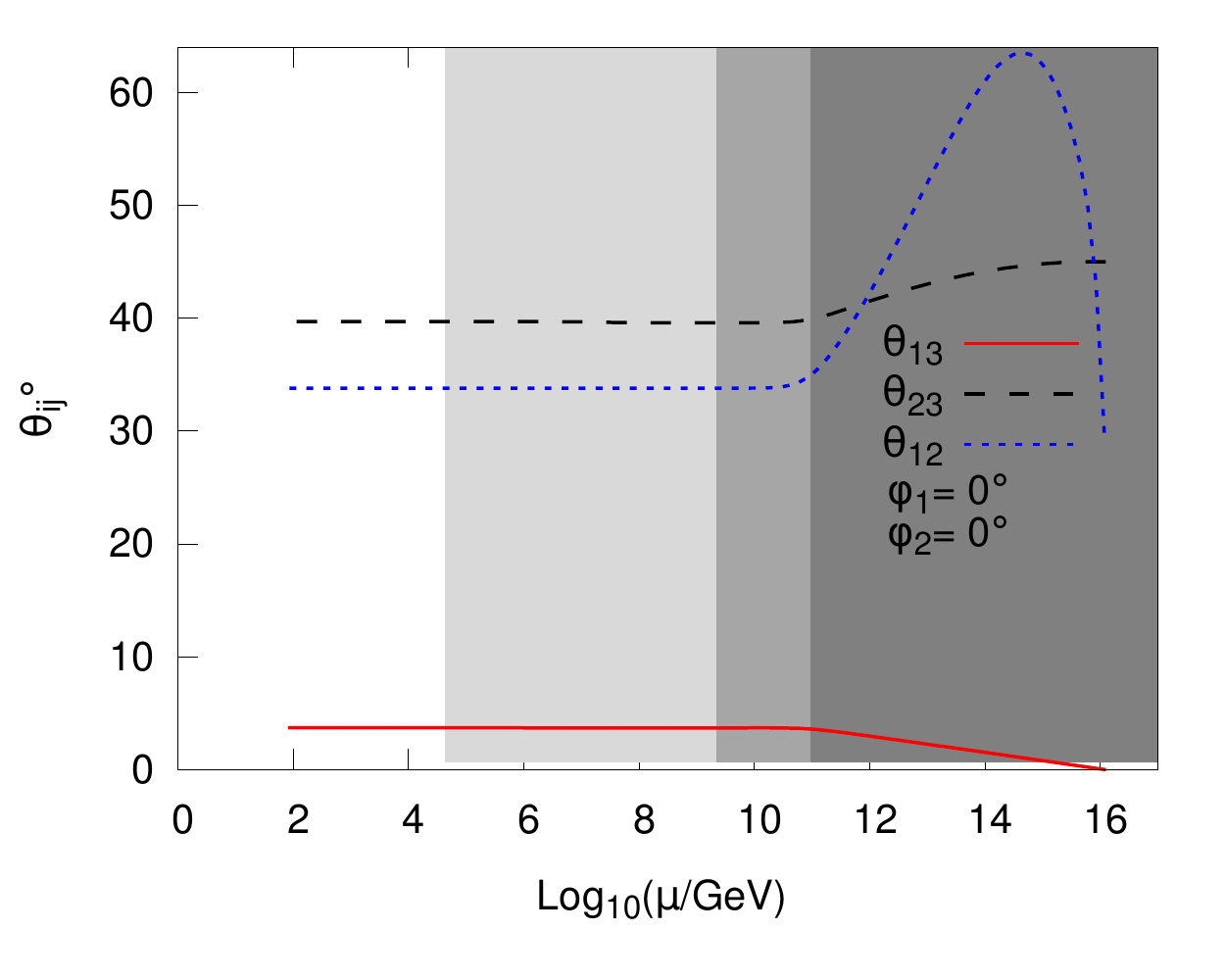}
\includegraphics[width=0.42\textwidth]{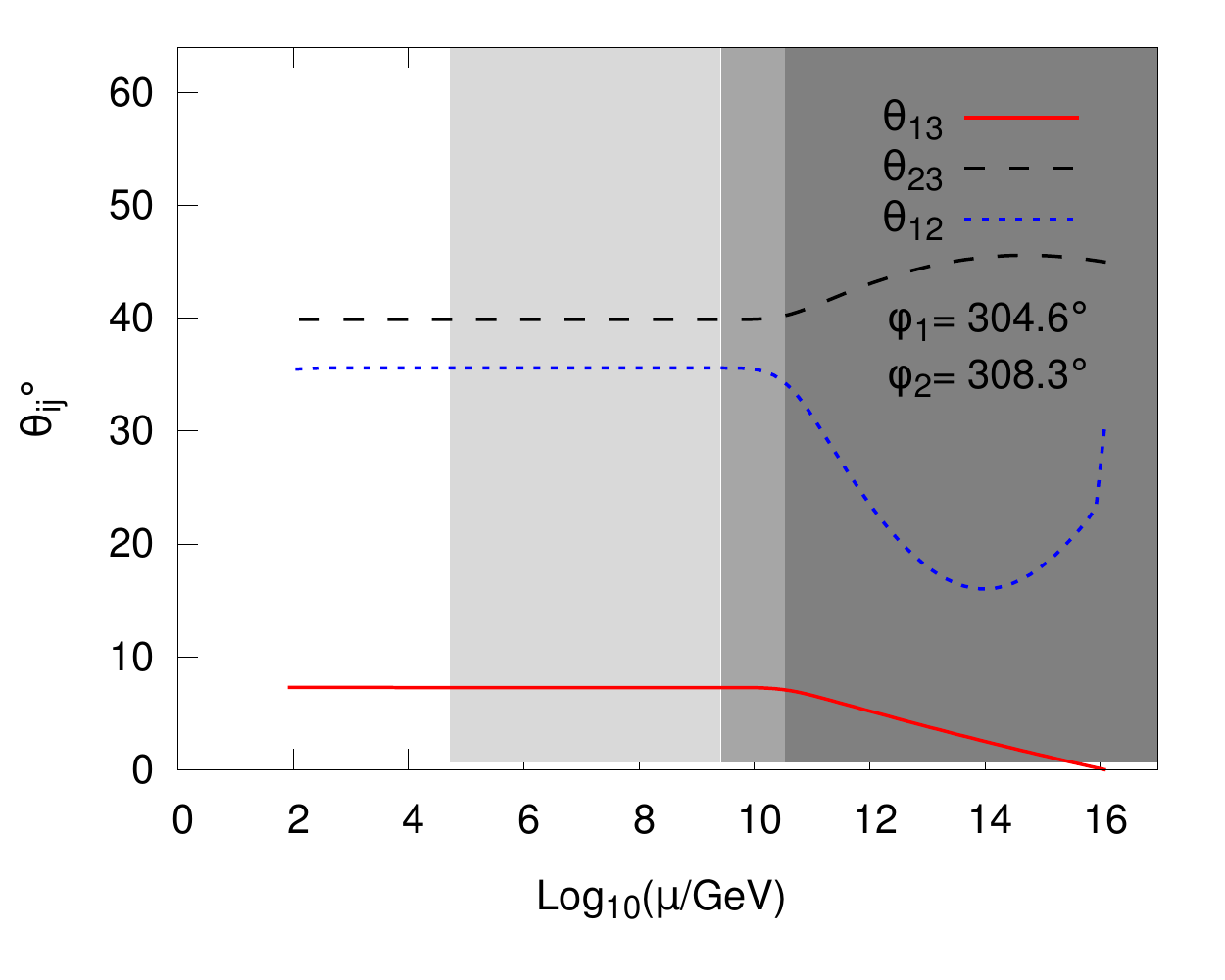}
\end{center}
 \begin{center}
\caption{The RGE  of the mixing angles between $\Lambda_{\rm GUT}$ and
 $\Lambda_{\rm EW}$ in the MSSM with tan$\beta$=10 for HM mixing. The input
parameters are given in the second and third column of Table 9. The grey shaded areas illustrate the ranges of effective theories
when heavy right handed singlets are integrated out.}
 \end{center}
\end{figure}
\begin{table}[h]
\begin{small}
\label{resultsgr11}
\begin{center}
\begin{tabular}{c|cc|c|cc}
\hline
\hline
MSSM Input  & $\varphi_1,\varphi_2=0$  & $\varphi_1,\varphi_2\neq0$ & MSSM Output & $\varphi_1,\varphi_2=0$ &  $\varphi_1,\varphi_2\neq0$  \\
\hline
$r_1$ & 0.58$\times 10^{-3}$ & 0.5$\times 10^{-3}$& $M_{R_1}$ (GeV) & 7.4$\times$10$^4$  & 3.87$\times$10$^4$ \\
$r_2$ & 0.68 & 0.53&$M_{R_2}$ (GeV)& 4.6 $\times$10$^9$ & 3.6$\times$10$^9$\\
$\delta$ & 349.5$^\circ $ & 235.5$^\circ$&$M_{R_3}$ (GeV) & 1.36 $\times$ 10$^{11}$ & 3.8$\times$ 10$^{10}$\\
$y_{\nu}$& 0.67 &  0.7& -& -&-\\
$\theta_1$ & 242.4$^\circ$  &169$^\circ$ &$\theta_{12}$ &35.3$^\circ$  & 34.8$^\circ$  \\
$\theta_2$ & 116.3$^\circ$ &233.2$^\circ$ &$\theta_{23}$ & 39.1$^\circ$ & 41.3$^\circ$ \\
$\theta_3$ & 105$^\circ$ & 46.2$^\circ$  &$\theta_{13}$ & 3.7$^\circ$ & 8$^\circ$\\
$ m_1$ (eV) & 2.1$ \times 10^{-3}$ & 4.7$ \times 10^{-3}$ &$ m_1$ (eV) & 6.5$ \times 10^{-4}$ & 2.8$ \times 10^{-3}$\\
$\Delta m^2_{12}$(eV$^2$) & 8.8 $\times$ 10$^{-7}$ & 4.7 $\times$ 10$^{-7}$ & $\Delta m^2_{12}$(eV$^2$) & 7.05 $\times$ 10$^{-5}$ & 7.2 $\times$ 10$^{-5}$\\
$\Delta m^2_{13}$(eV$^2$) & 3.15$\times 10^{-3}$ & 3.0$\times 10^{-3}$&$\Delta m^2_{13}$(eV$^2$) & 2.47$\times$ 10$^{-3}$ & 2.6$\times$ 10$^{-3}$ \\
$\varphi_1$& 0$^\circ$ & 161.2$^\circ$&$\varphi_1$ & 319.5$^\circ$ &230$^\circ$\\
$\varphi_2$& 0$^\circ$ & 325$^\circ$&$\varphi_2$ & 302.7$^\circ$ & 92.8$^\circ$\\
--&--&--&$J_{CP}$ &  -5.6$\times$10$^{-3}$ & 0.012 \\
--&--& --&$M_{ee} (eV)$ & 3.0$\times$10$^{-3}$ & 2.6$\times$10$^{-3}$ \\
\hline
\end{tabular}
\end{center}
\begin{center}
\caption{Numerical values of input and output parameters radiatively generated in the MSSM for GR mixing
for zero and non zero Majorana phases at $\Lambda_{\rm GUT}$ = 2$\times 10^{16}$GeV and tan$\beta$=10.
The input values for neutrino mixing angles at GUT scale are $\theta_{13}$= 0$^\circ$, $\theta_{23}$= 45$^\circ$ and $\theta_{12}$=31.7 $^\circ$.}
 \end{center}
 \end{small}
\end{table}
\begin{figure}[h]
\begin{center}
\label{figgrmssm}
\includegraphics[width=0.42\textwidth]{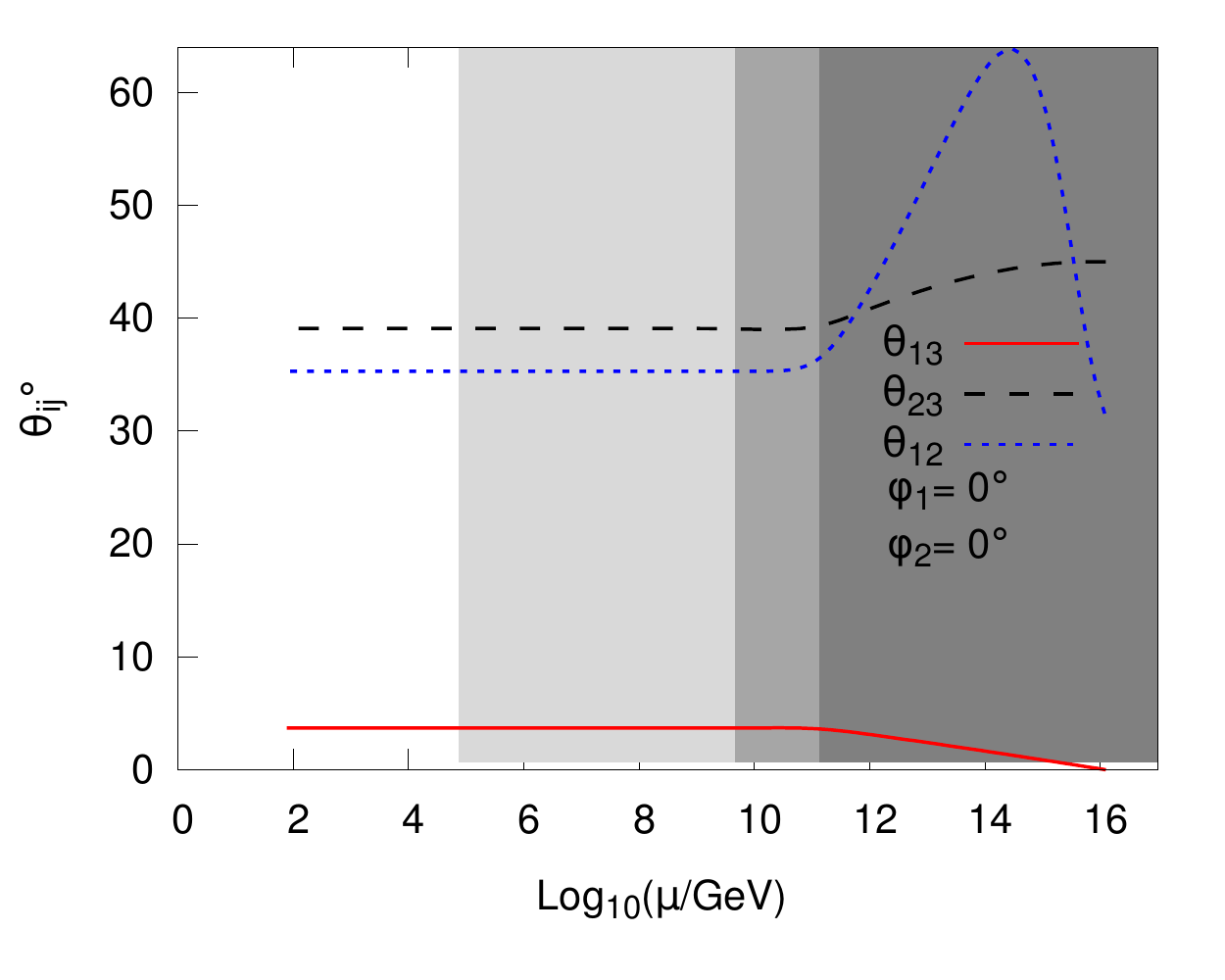}
\includegraphics[width=0.42\textwidth]{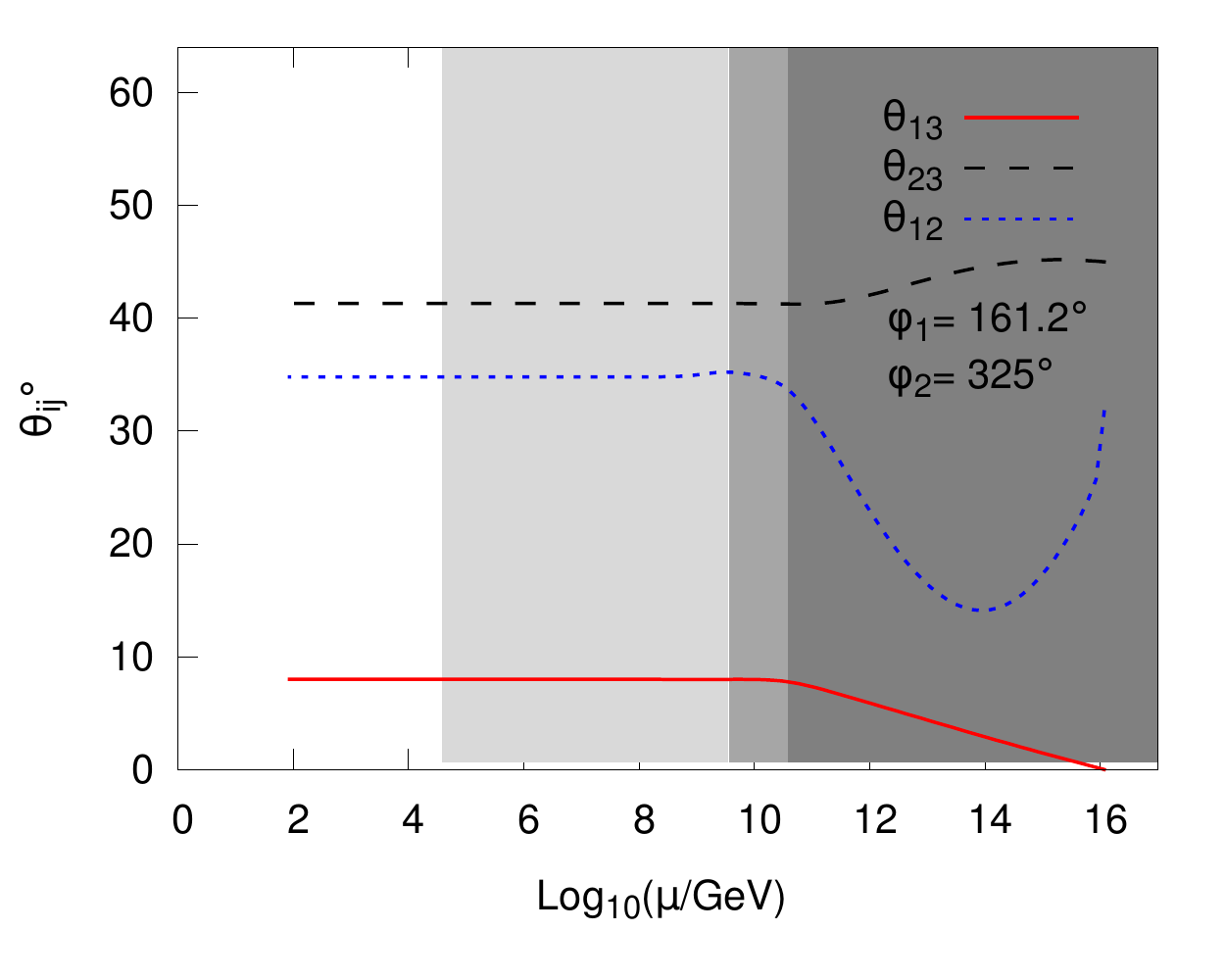}
\end{center}
 \begin{center}
\caption{The RGE  of the mixing angles between $\Lambda_{\rm GUT}$ and $\Lambda_{\rm EW}$
 in the MSSM with tan$\beta$=10 for GR mixing.
The initial values of the parameters are given in the third column of Table 10. The grey shaded areas illustrate the
ranges of effective theories when heavy right handed singlets are integrated out.}
 \end{center}
\end{figure}
\subsection{RGE and seesaw threshold corrections in the MSSM}
The study of radiative corrections in the MSSM in presence of seesaw threshold effects can again
be divided into three regions as in the SM. All these regions will be governed by different RGE equations, respectively.
The RGE corrections to the mixing angles in the MSSM for region below seesaw scales can be larger than the SM
due to the presence of factor $Y_{\tau}^2$ (1+tan$^2\beta$).
This term can be large when tan$\beta$ is large, thus, resulting in significant
changes in the mixing angles where $Y_l$ is the only contributing term.
We study the RGE of mixing angles with zero and non zero Majorana phases in the MSSM with tan$\beta$=10.

As seen in Figs 5, 6, 7 and 8, the RGE running effects are small below lowest seesaw scale $M_{R_1}$ down to the EW scale for tan$\beta=$10.
In the region above seesaw scale, $M_{R_3}$ there is large running of all the mixing angles as can be seen
from the figures of all mixing scenarios in the MSSM. This is due to the contribution
of $Y_{\nu}$ which can be large regardless to value of tan$\beta$
 in addition of $Y_l$. We have the large running of all mixing angles including
 $\theta_{13}$ in this region when Majorana phases are nonzero.
 Between the seesaw scale there is very small running in comparison to the SM.
This behavior is described in Ref. \cite{Antusch:2005gp} in detail as the enhanced running
 between the threshold scales due to the term with trivial flavor structure is absent in the MSSM.
For vanishing Majorana phases , $\theta_{13}<$ 3.7$^\circ$
is obtained in all mixing scenarios. However, when Majorana phases are considered the largest possible value of $\theta_{13} \approx$ 9.46
$^\circ$ is obtained in the TBM mixing. We get $\theta_{13}$ within its allowed 3$\sigma$ range at the low scale in all scenarios.
The Majorana phases play an important role in the enhancement of
RGE and thus, $\theta_{13}$ can be produced in its 3$\sigma$ allowed range
at the EW scale in all mixing scenarios along with the other neutrino oscillation parameters.
Three seesaw scales $M_{R_i}$, given in output of all tables are determined from Eqs.(\ref{Mr1}, \ref{Mr2}).
For the MSSM, in TBM when Majorana phases are zero we get $M_{R_i}$ as 9.85$\times$10$^3$eV,  2.2$\times$10$^{9}$eV,  4.3$\times$10$^{10}$eV at the GUT scale using Eq.(\ref{Mr1}).
The different seesaw threshold scales in this case are given in Table 7. There is small difference in the values due to running between the GUT and seesaw scales. Running of $M_{r_1}$ towards higher value is observed here.

 We also see the radiative corrections to the masses in Fig. 5.
The running of masses, however, as in the SM is independent of the mixing parameters
since $\alpha$ is usually much larger than $Y_{\tau}^2$ (1+tan$^2\beta$)
except in the MSSM with large tan$\beta$. RGE effects of neutrino masses are smallest if tan$\beta$=10.
The negligible running of masses is seen below the seesaw scales irrespective of values of
$\varphi_1$ and $\varphi_2$ which indicate that the running of
masses is not directly dependent on the Majorana phases \cite{casas2}. From RGEs of mixing angles and the Dirac phase $\delta_{CP}$ which depend on the Majorana phases,
we obtain J$_{CP}$ $\approx$10$^{-2}$ and the effective
Majorana mass $M_{ee} \approx$10$^{-3}$eV at low scale for all mixing scenarios.
Thus, large value of $\theta_{13}$ which is in its present 3$\sigma$ range at EW
scale can be produced in the MSSM for tan$\beta$=10 when $\varphi_1$ and $\varphi_2$ are both non zero at high scale.

\section{Conclusions}

We assume different lepton mixing matrices at the high energy (GUT) scale and study
effects of the RGE  and seesaw threshold corrections
to these mixing scenarios both in the SM and MSSM. In the absence of seesaw threshold effects
there are very small corrections both in the SM and MSSM. 
Significant corrections are observed both in the SM and  MSSM when threshold effects are included.
Above the seesaw scales there are more number of parameters due to $Y_{\nu}$ that can significantly
 affect the RGE of mixing angles. Below the lowest seesaw scale, contribution of $Y_{\nu}$ is absent and the RGE corrections are only due to $Y_l$ which is very small in the SM and MSSM with small tan$\beta$. For large tan$\beta$ however,
 there can be significant contribution below the lowest seesaw scale in the MSSM.
Some of these mixing scenarios are studied in \cite{xing} at high scale without fully considering the effects of Majorana CP phases.
In that case our results are somewhat similar and $\theta_{13}<$ 5$^\circ$ is obtained at low energy.
The Majorana phases, however, play a significant role in the running of parameters.
When non zero value of Majorana phases are considered at the high scale,
it is possible to enhance $\theta_{13}$ to its allowed 3$\sigma$ range in the MSSM. 
Here, we presented a comprehensive study by considering four different mixing scenarios
 at the GUT scale and study their running behavior in the SM and MSSM with tan$\beta$=10.
We conclude that for TBM, BM, HM and GR mixings at some high scale, say the GUT scale,
the RGE and seesaw threshold corrections can result in significant corrections to the mixing angles both in the SM and MSSM at the low energy scale.  In the MSSM with tan$\beta$=10 it is possible to simultaneously obtain all neutrino mixing angles and mass squared
differences in their present 3$\sigma$ ranges at the EW scale
when the Majorana phases are considered.
Finally we note that the input values of $M_{R_1}$ taken in our numerical analysis are too small to achieve the successful leptogenesis
via the decay of the lightest heavy Majorana neutrino, so a variation of the leptogenesis is required, which is beyond the scope of this work.

\section{Acknowledgements}
The work of C.S.K and S.G. is supported by the National Research Foundation of Korea (NRF) grant funded by
Korea government of the Ministry of Education, Science and Technology (MEST) (Grant No. 2011-0017430 and Grant No. 2011-0020333).

\end{document}